\documentclass[aps,twocolumn,pra,bibliography]{revtex4-2}
\usepackage[utf8]{inputenc}
\usepackage[T1]{fontenc}
\usepackage{amsmath}

\usepackage{physics}
\usepackage{mathtools}
\usepackage{amsfonts}
\usepackage{amssymb}
\usepackage{graphicx}
\usepackage{caption}
\usepackage{ragged2e}
\usepackage{subcaption}
\usepackage{textcomp}
\usepackage{color}
\usepackage[dvipsnames]{xcolor}
\usepackage{placeins}

\begin{document}

\title{Simulating a quasiparticle on a quantum device}
\author{Rimika Jaiswal}
\affiliation{Department of Physics, University of California Santa Barbara, CA}
\author{Izabella Lovas}
\affiliation{Kavli Institute for Theoretical Physics, Santa Barbara, CA}
\author{Leon Balents}
\affiliation{Kavli Institute for Theoretical Physics, Santa Barbara, CA}
\affiliation{Canadian Institute for Advanced Research, Toronto, Ontario, Canada}

\date{\today}

\begin{abstract}
We propose a variational approach to explore quasiparticle excitations in interacting quantum many-body systems, motivated by the potential in leveraging near-term noisy intermediate scale quantum devices for quantum state preparation. By exploiting translation invariance and potentially other abelian symmetries of the many-body Hamiltonian, we extend the variational quantum eigensolver (VQE) approach to construct spatially localized quasiparticle states that encode information on the whole excited band, allowing us to achieve quantum parallelism. We benchmark the proposed algorithm via numerical simulations performed on the one-dimension transverse field Ising chain. Our numerical results demonstrate that VQE can capture both the magnon quasiparticles of the paramagnetic phase, and the topologically non-trivial domain wall excitations in the ferromagnetic regime. We argue that the localized quasiparticle states constructed with VQE contain accessible information on the full band of quasiparticles, and provide valuable insight into the way interactions renormalize the bare spin flip or domain wall excitations of the simple, trivially solvable limits of the model. These results serve as important theoretical input towards utilizing quantum simulators to directly access the quasiparticles of strongly interacting quantum systems, as well as to gain insight into crucial experimentally measured properties directly determined by the nature of these quasiparticles.
\end{abstract}

\maketitle

\section{Introduction}

Recent advances in experimental techniques have opened unprecedented possibilities to study quantum many-body systems in quantum simulators realized on various platforms, such as ultracold atoms~\cite{Bloch_RevModPhys.80.885,Bernien2017},
trapped ions~\cite{Blatt2012,trappedion_doi:10.1126/science.1232296} and superconducting qubits~\cite{SCqubit_PhysRevLett.111.080502,Gambetta2017}. Complex quantum many-body states prepared in these settings, hosting non-trivial correlations and entanglement patterns, have the potential to serve as valuable resources for quantum computation and precision measurements in quantum metrology~\cite{Giovannetti2011,RevModPhys.89.035002}. Furthermore, gaining direct access to complicated quantum many-body states grants invaluable insights into the phases and correlated structure of strongly interacting quantum systems, and may unveil properties beyond the reach of simulations on classical computers.

Despite recent successes, efficient quantum state preparation, benefiting from computational advantages over classical simulations, remains a challenging task, hindered by the lack of large scale fault-tolerant quantum computers. This drives intense interest in leveraging currently available or near-term noisy intermediate-scale quantum devices~\cite{Preskill2018quantumcomputingin}, by proposing generic, versatile protocols that can run on these architectures. Variational quantum-classical simulations are receiving particular attention as promising applications of near-term quantum devices~\cite{Peruzzo2014,VQE_PhysRevX.6.031045,Kandala2017,SciPostPhys.6.3.029,TILLY20221,excited_PhysRevResearch.1.033062,thermal_PhysRevLett.123.220502, McClean_2016, Higgott2019variationalquantum}. These hybrid quantum-classical computational approaches rely on a feedback loop, where the state preparation and a measurement subroutine happen on a quantum computer while a classical computer is used to process the measurement results and update the quantum computer according to an update rule. This formalism is well suited for constructing Variational Quantum Eigensolvers (VQEs) designed to prepare the ground states of Hamiltonians~\cite{TILLY20221,VQE_PhysRevX.6.031045,Peruzzo2014,Kandala2017,SciPostPhys.6.3.029}. 

While the focus in most of these simulations is on preparing novel ground states, in strongly interacting quantum systems often the most important properties are understood in terms of \emph{quasiparticles}~\cite{Abrikosov:107441,NegeleOrland,Coleman_2015}.  Quasiparticles are elementary excitations of the system which behave like particles.  Even in a ``trivial" ground state, i.e. one which is smoothly connected to a simple state like a product state, the quasiparticles can be strongly dressed and highly renormalized from any simple soluble limit.  The canonical example of tremendous importance in condensed matter physics is the electronic quasiparticle in a metal, posited originally by Landau.  Landau's Fermi liquid theory explains that electronic quasiparticles are described by a renormalized dispersion near the Fermi energy (e.g. effective mass), a quasiparticle weight $Z$, and Fermi liquid interaction parameters~\cite{Abrikosov:107441,NegeleOrland,Coleman_2015}. These properties are difficult to calculate in general, but for example, can be approximately obtained using dynamical mean field theory, and can be measured for example by photoemission spectroscopy and other means.  

Given the power of quantum simulators, it is intriguing to ask if quasiparticle states and properties can be studied therein.  The notion is not limited to Fermi liquids but exists for any many body ground state away from a quantum critical point (where there need not be any quasiparticle basis \emph{a priori}).  The nature of the quasiparticle depends upon the system and its ground state, and may even be emergent, i.e. the quasiparticle may not be adiabatically connected to any weakly interacting eigenstate or any state that can be created by the action of a local operator.  Irrespective of whether quasiparticles are emergent or not, a quantitative understanding of their dispersion and other properties is important because it directly predicts many experimentally measured properties such as dynamical correlation functions, transport coefficients, etc.

In this work, we address a potential application of quantum devices in studying quasiparticles, by relying on the framework of VQE. We focus on the simplest case of the transverse field Ising chain, which supports different types of quasiparticles in its paramagnetic and ferromagnetic phases~\cite{pfeuty1970}. In the paramagnetic phase, it hosts magnon quasiparticles, localized excitations that can be understood as single spin flips dressed by interactions, whereas in the ferromagnetic phase, the nature of these excitations changes to domain wall (soliton) quasiparticles\cite{Mikeska1996}. By combining theoretical arguments and numerical simulations, we demonstrate that VQE can effectively produce these localized excitations in both phases, even when they are highly renormalized from the non-interacting limit. These so-called “Wannier states” of quasiparticles contain information on all momentum eigenstates forming an excited band, therefore, the proposed VQE protocol achieves quantum parallelism.  We discuss how to extract detailed information on the excited band from these simulations, including the band gap and the full band dispersion. We also explore the properties of Wannier states obtained in our numerical simulations of the VQE protocol, by examining the width of localized quasiparticles, as well as a quasiparticle weight quantifying the renormalization compared to their non-interacting counterparts.

The VQE protocol proposed in this paper relies heavily on the symmetries of the Hamiltonian, most prominently on translation invariance.   The potential of leveraging symmetries to enhance the efficiency of VQE has been discussed in prior works, particularly in the context of quantum chemistry~\cite{PhysRevA.99.062304, Gard2020, Ryabinkin2019}, where targeting and monitoring variational states in a given symmetry sector allows one to mitigate the errors on noisy devices~\cite{PhysRevA.99.062304} or to probe the energies of molecular states in different charge or spin sectors~\cite{Gard2020, Ryabinkin2019}. Related ideas have also been applied to explore the low energy excitations in many-body systems, by imprinting the desired symmetry eigenvalue on the variational state via non-unitary projectors realized as part of a classical postprocessing step ~\cite{symm_adapt_PhysRevA.101.052340}. Our work differs from these earlier approaches by introducing a comprehensive framework explicitly focusing on Wannier states of quasiparticles instead of individual momentum eigenstates. As we discuss in subsequent sections, the resulting protocol offers considerable computational advantages, most importantly, it facilitates to benefit from  quantum parallelism — a pivotal advantage of quantum simulators. Furthermore, directly targeting localized quasiparticle states can also be advantageous from a theoretical standpoint, since these localized elementary excitations can serve as essential building blocks for theoretical considerations.

The rest of the paper is organized as follows. We present the theoretical framework applied in this work in Sec.~\ref{sec:background}, with Sec.~\ref{subseq:tfim} devoted to a brief review of a few basic properties of the transverse field Ising model relevant for our discussion, and Sec.~\ref{subsec:vqe} turning to the VQE approach and its application for preparing quasiparticle excitations,  guided by the symmetries of the Hamiltonian. We then benchmark the proposed VQE protocol by performing numerical simulations on the transverse field Ising model. We first focus on the magnon quasiparticles of the paramagnetic phase in Sec.~\ref{sec:magnon}, then consider the soliton quasiparticles of the ferromagnetic regime in Sec.~\ref{sec:dw}. We conclude by discussing potential extensions and the future scope of our work in Sec.~\ref{sec:outlook}. We present supplementary numerical results from our VQE simulations in the Appendices.

\section{Theoretical background}\label{sec:background}

In this paper, we extend the framework of VQE to study quasiparticles. While the approach is more general, we focus on the paradigmatic example of the one-dimensional Transverse Field Ising Model (TFIM), an ideal test bed for exploring different types of quasiparticles. Besides serving as a convenient starting point for theoretical and numerical considerations, the TFIM also has fundamental practical relevance. Although real-world materials often exhibit more complex behaviors, many fall within the same universality class as the TFIM~\cite{Sachdev1999}. Furthermore, the TFIM effectively models the low-energy properties of certain quasi-1D materials, such as Cobalt niobate, to the lowest order~\cite{Coldea2010}. We also note that the highly entangled Greenberger-Horne-Zeilinger (GHZ) states emerging in the ferromagnetic phase of the TFIM receive a lot of attention as valuable resources for quantum information processing and quantum metrology~\cite{Leibfried2004,Giovannetti2011,RevModPhys.89.035002}.

Below we first briefly review a few basic properties of the TFIM in Sec.~\ref{subseq:tfim}. We then turn to the framework of VQE in Sec.~\ref{subsec:vqe}, putting a particular emphasis on the role of symmetries of the Hamiltonian.

\subsection{Review of TFIM}\label{subseq:tfim}

We consider the one-dimensional TFIM, describing a chain of  interacting spin-$\frac{1}{2}$ particles with the  Hamiltonian given by
\begin{equation}
    H_{\rm TFIM} = - J \sum_{i=1}^{N} Z_i Z_{i+1} - h \sum_{i=1}^{N} X_i = - J H_{ZZ} - h H_{X}.
    \label{eq:ham}
\end{equation}
Here $X_i, Z_i$ denote the Pauli spin operators at site $i$,  $N$ is the system size, and we took periodic boundary conditions with the site indices $i$ understood modulo $N$. The Hamiltonian comprises of two non-commuting parts introducing complex dynamics into the system: $H_{ZZ}$, which represents the Ising interaction between adjacent spins in the $Z$ direction, and $H_{X}$, accounting for a transverse magnetic field in the $x$ direction. The competition between $H_{ZZ}$ and $H_{X}$ leads to quantum fluctuations that drive the system through a quantum phase transition between a ferromagnetic and a paramagnetic phase at a critical value of the transverse field $h/J =1$. Furthermore, these terms are responsible for generating different types of renormalized excitations, crucial for understanding the low-energy physics of the model.

Specifically, two distinct types of quasiparticles emerge whose properties can be intuitively understood in the two separate limits of the model: the high magnetic field limit $J/h \to 0$ hosting spin flip excitations, and the high coupling limit $J/h \to \infty$ with domain wall excitations. In the paramagnetic regime $J/h<1$,  spin-flip like quasiparticles called magnons emerge due to the strong magnetic field. Conversely, in the ferromagnetic limit $J/h>1$ where the Ising coupling dominates, the primary excitations are domain walls or solitons, which represent transitions between different ground state configurations across the chain. Below we will first discuss the characteristics and preparation of magnon quasiparticles in Sec.~\ref{sec:magnon} and subsequently explore the dynamics of soliton quasiparticles in Sec.~\ref{sec:dw}.

We conclude this section by commenting on the symmetries of the Hamiltonian. These symmetries, notably parity and translation, will play a prominent role in the discussion that follows. Parity symmetry is represented by $P = \prod_{i} X_{i}$, flipping all the spins across the lattice, and has eigenvalues $p=\pm 1$. Translation symmetry, denoted by $T$, translates each spin by one site along the chain and has eigenvalues $e^{ik}$ with $k$ labeling the first Brillouin Zone. Mathematically, these symmetries ensure that the Hamiltonian remains unchanged under their operations, expressed as
\begin{equation}
    T^\dagger H T = H
    \quad \text{and} \quad
    P^\dagger H P = H.
\end{equation}
The eigenstates of the TFIM can thus be labeled by the parity and momentum quantum numbers, $p$ and $k$ such that
\begin{equation}
    P \ket{\psi_{p,k}} = p \ket{\psi_{p,k}}
    \quad \text{and} \quad
    T \ket{\psi_{p,k}} = e^{ik} \ket{\psi_{p,k}},
\end{equation}
a fact we will exploit in subsequent numerical simulations. We note that we will continue to use the notation $P$ for the symmetry operator, and $p$ for a particular symmetry eigenvalue throughout the paper.

\begin{figure}[h!]
    \centering
    \includegraphics[width=\columnwidth]{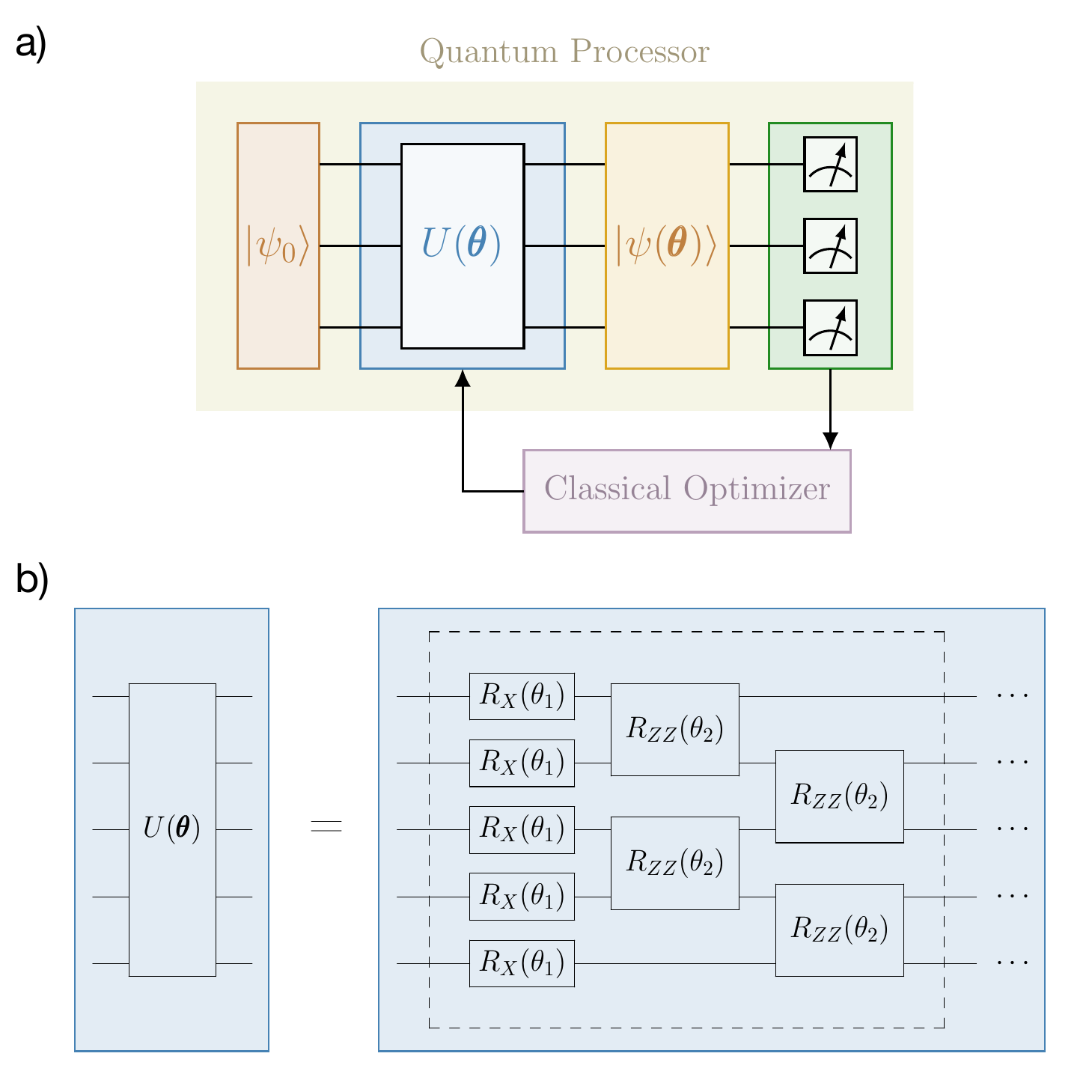}
    \caption{\justifying{
    \textbf{Symmetry-guided VQE framework.} (a) Schematic illustration of the hybrid quantum-classical VQE approach, preparing a quantum many-body state on a quantum processor variationally. The initial wave function $\ket{\psi_0}$ is an easy-to-prepare short range entangled state within the same symmetry sector of the Hilbert space as the target state. A trial state $\ket{\psi({\pmb\theta})}$ is prepared on a quantum processor by applying a symmetry preserving parametrized unitary gate  $U(\pmb \theta)$. One then performs measurements on $\ket{\psi({\pmb\theta})}$, and feeds the measurement outcomes to a classical optimizer, updating the parameters $\pmb\theta$ to minimize a suitable cost function, commonly chosen as the expectation value of a many-body Hamiltonian. These steps are repeated until convergence.
    (b) Structure of unitary operator $U(\pmb \theta)$, used in VQE to prepare quasiparticles of the TFIM. We use a parametrized unitary circuit consisting of $d$ blocks, built from alternating layers representing the non-commuting terms of $H_{\rm TFIM}$, $H_{X}$ and $H_{ZZ}$, through the single qubit rotations $R_X(\theta)\equiv e^{-i \theta X_j}$ and entangling gates of nearest-neighbor pairs $R_{ZZ}(\theta)\equiv e^{-i \theta Z_j Z_{j+1}}$.  The parameters $\{\theta_1,...,\theta_{2d}\}$ appearing in this ansatz are optimized during the VQE process.}}
    \label{fig:VQE_combined}
\end{figure}

\subsection{Symmetry-Guided VQE}\label{subsec:vqe}

We now turn to the VQE scheme applied in this paper, making use of the symmetries of the Hamiltonian to gain information on a full excited band in a parallelized way. We first introduce the general formalism, and then discuss the specific algorithm applied for the TFIM.

\subsubsection{General framework}

The objective of the VQE approach is to prepare a close approximation of a target wavefunction, $\ket{\psi_T}$, from an easy-to-prepare initial state $\ket{\psi^0}$. This is achieved by constructing an approximation for the unitary transformation $U_T$ that  \textit{exactly} transforms $\ket{\psi^0}$  into the target state $\ket{\psi_T}$, $\ket{\psi_T} = U_T \ket{\psi^0}$, via the variational method illustrated in Fig.~\ref{fig:VQE_combined}a ~\cite{VQE_PhysRevX.6.031045,Peruzzo2014,Kandala2017,SciPostPhys.6.3.029,TILLY20221}. Here, one chooses a parameterized unitary circuit Ansatz $U(\pmb \theta)$, and optimizes the parameters $\pmb{\theta}$ by minimizing a cost function $C(\pmb \theta)$. The form of the Ansatz $U(\pmb \theta)$ and the cost function are chosen to ensure that $\ket{\psi(\pmb{\theta})} = U(\pmb \theta) \ket{\psi^0}$ converges towards the desired target state as the number of variational parameters increases, $ \norm{\ \ket{\psi(\pmb \theta)} - \ket{\psi_T}} \to 0$. In the important example when the target $\ket{\psi_T}$ is the ground state of a Hamiltonian $H$, a natural choice for the cost function is the expectation value of the Hamiltonian in the variational state,  $C(\pmb \theta)\equiv E_{\pmb \theta}\equiv\expval{H}{\psi(\pmb \theta)}$. As mentioned in the introduction, the resulting protocol is one of the promising applications of near-term quantum devices, where the preparation of the state $\ket{\psi(\pmb{\theta})}$ would run on a quantum simulator, followed by the measurement of $H$. An updated parameter set $\pmb{\theta}$ for the next iteration would then be calculated on a classical computer, closing the hybrid quantum-classical feedback loop (see Fig.~\ref{fig:VQE_combined}a).

We utilize symmetries to extend this method for preparing quasiparticles in translation invariant systems. Suppose we aim to prepare an excited state of a Hamiltonian $H$, exhibiting a set of symmetries $\{G_1, G_2, \cdots, G_n\}$, satisfying $[H, G_i]= 0$ for all $i$. We make the further assumption that the symmetries are abelian, $[G_i,G_j]=0$ for all $i,j$. Here we are keeping the discussion general, but later we will focus on settings with translation symmetry $T$, and choose $G_1\equiv T$. The eigenstates of $H$ can be labeled by a set of quantum numbers which are eigenvalues of these symmetry operators, $\mathbf{\Lambda}=\{\lambda_{G_1}, \lambda_{G_2} \cdots, \lambda_{G_n}\}$. The first step in preparing the target excited state is identifying the symmetry sector it belongs to within the full Hilbert space $\mathcal{H}$, $\ket{\psi_T}\in \mathcal{H}_{\mathbf{\Lambda}}  \subset \mathcal{H}$, with  $\mathcal{H}_{\mathbf{\Lambda}}$ denoting the symmetric subspace specified by the quantum numbers $\mathbf{\Lambda}$.

In the VQE approach, it is convenient to choose the parametrized quantum circuit $U(\pmb \theta)$ in such a way that it preserves the symmetries of the Hamiltonian~\footnote{Full translation symmetry is broken by random circuits consisting of local unitary gates, but they can still preserve symmetry against translation by two lattice sites.}, ensuring that the trial state remains within the same symmetry subspace as the initial state $\ket{\psi^0}$,
\begin{equation}
\ket{\psi(\pmb \theta)} = U(\pmb \theta) \ket{\psi^0} \in \mathcal{H}_{\mathbf{\Lambda}} \quad \text{if} \quad \ket{\psi^0} \in \mathcal{H}_{\mathbf{\Lambda}}.
\end{equation}
This observation can be exploited to prepare low energy excitations with VQE that belong to a symmetry sector different from the ground state, and correspond to the lowest energy in that subspace. We note that an alternative symmetry-adapted VQE algorithm has been put forward in \cite{symm_adapt_PhysRevA.101.052340}, proposing to imprint the desired symmetry eigenvalues by applying projectors to the non-symmetric output states $\ket{\psi(\pmb \theta)}$. While this work has introduced ideas related to our approach, it has not focused on Wannier states. As we show below, targeting Wannier states offers considerable advantages and allows us to obtain a general numerically efficient framework to utilize symmetries, which is especially well suited for achieving quantum parallelism.

An additional important requirement for practical applications is that the initial state $\ket{\psi^0}$ should be sufficiently simple, i.e., it has to be a product state or a state that can be prepared with a low-depth unitary circuit. This can be easily satisfied for symmetries that are written as products of on-site operators, such the parity symmetry in the TFIM acting as a spin flip performed on each site, but is not true for momentum eigenstates apart from the uniform state $k=0$. While choosing $\ket{\psi^0}$ as a momentum eigenstate is impractical, we can still make use translation invariance by noting that a  unitary transformation $U(\pmb \theta)$  preserving the translation symmetry also conserves the weight of different momentum components in $\ket{\psi(\pmb \theta)}$. We instead use a product initial state that is an equal weight superposition of different momentum eigenstates,
\begin{equation}
\ket{\psi^0} = \dfrac{1}{\sqrt{N}}\sum_{k} \ket{\psi_k^0},
\end{equation}
where $T \ket{\psi_k^0}=e^{ik}\ket{\psi_k^0}$, and the sum runs over the first Brillouin zone. As detailed below, in the case of the TFIM $\ket{\psi^0}$ will correspond to a single spin flip on a background polarized into the $X$ direction, and a single domain wall in the paramagnetic and ferromagnetic phases, respectively. This choice ensures that $\ket{\psi(\pmb \theta)}$ will also be an equal weight superposition of momentum eigenstates $\ket{\psi_k(\pmb \theta)}$, thereby  describing  a localized excitation, in other words a Wannier state of quasiparticles. Importantly, the expectation value of the Hamiltonian can still serve as the cost function in this procedure, since
\begin{equation}\label{eq:cost}
   E_{\pmb{\theta}}= \expval{H}{\psi(\pmb \theta)}=\dfrac{1}{N}\sum_k \expval{H}{\psi_k(\pmb \theta)},
\end{equation}
minimized when each $\ket{\psi_k(\pmb \theta)}$ is the lowest energy eigenstate with momentum $k$. Therefore, the state $\ket{\psi(\pmb \theta^*)}$ obtained with at the optimal parameters $\pmb{\theta}^*$ carries information on a whole excited band, allowing us to access different momentum eigenstates in a parallelized way. We will discuss the details of this protocol, as well as certain subtleties related to a phase freedom in the eigenstates $\ket{\psi_k(\pmb \theta)}$, in the subsequent sections, focusing on the TFIM.

\subsubsection{Application to TFIM}

Here, we demonstrate how to apply the symmetry-guided VQE framework to the TFIM informed by its parity symmetry and translation invariance. As mentioned above, it is important to choose a circuit ansatz that preserves both momentum and parity quantum numbers. We employ a circuit ansatz with alternating layers~\cite{SciPostPhys.6.3.029}, depicted in Fig. \ref{fig:VQE_combined}b, composed of the two non-commuting parts of the target Hamiltonian, $H_{ZZ}$ and $H_{X}$.  A circuit of depth $d$ includes sequential blocks, with the $i$th block consisting of a global spin rotation by angle $\theta_{2i-1}$ generated by $H_{X}$, followed by layers of $H_{ZZ}$ gates with parameter $\theta_{2i}$ connecting each nearest-neighbor pair of qubits,
\begin{equation}\label{eq:circuit}
    U(\pmb \theta) = e^{-i\theta_{2d} H_{ZZ}} e^{-i\theta_{2d-1} H_X} \cdots  e^{-i\theta_2 H_{ZZ}} e^{-i\theta_1 H_X}.
\end{equation}
The structure of the circuit, consisting of on-site rotations at each site $j$, $R_X(\theta)\equiv e^{-i \theta X_j}$, and entangling gates between neighboring sites $j$ and $j+1$, $R_{ZZ}(\theta)\equiv e^{-i \theta Z_j Z_{j+1}}$, is shown in Fig. \ref{fig:VQE_combined}b. We note that the same circuit ansatz was used in Ref.~\cite{SciPostPhys.6.3.029} to variationally prepare the ground state of the TFIM. As customary, we use the expectation value of the Hamiltonian \eqref{eq:ham} in the trial state $\expval{H_{\rm TFIM}}{\psi(\pmb \theta)}$ as the cost function. Furthermore, we always choose the initial state $\ket{\psi^0}$ as an eigenstate of an exactly solvable Hamiltonian  $H_{X}$ or  $H_{ZZ}$, depending on whether $H_{\rm TFIM}$ falls in paramagnetic or ferromagnetic phase, with the precise structure of  $\ket{\psi^0}$ depending on our target state.

\begin{figure*}[t!]
    \centering
    \includegraphics[width=\textwidth]{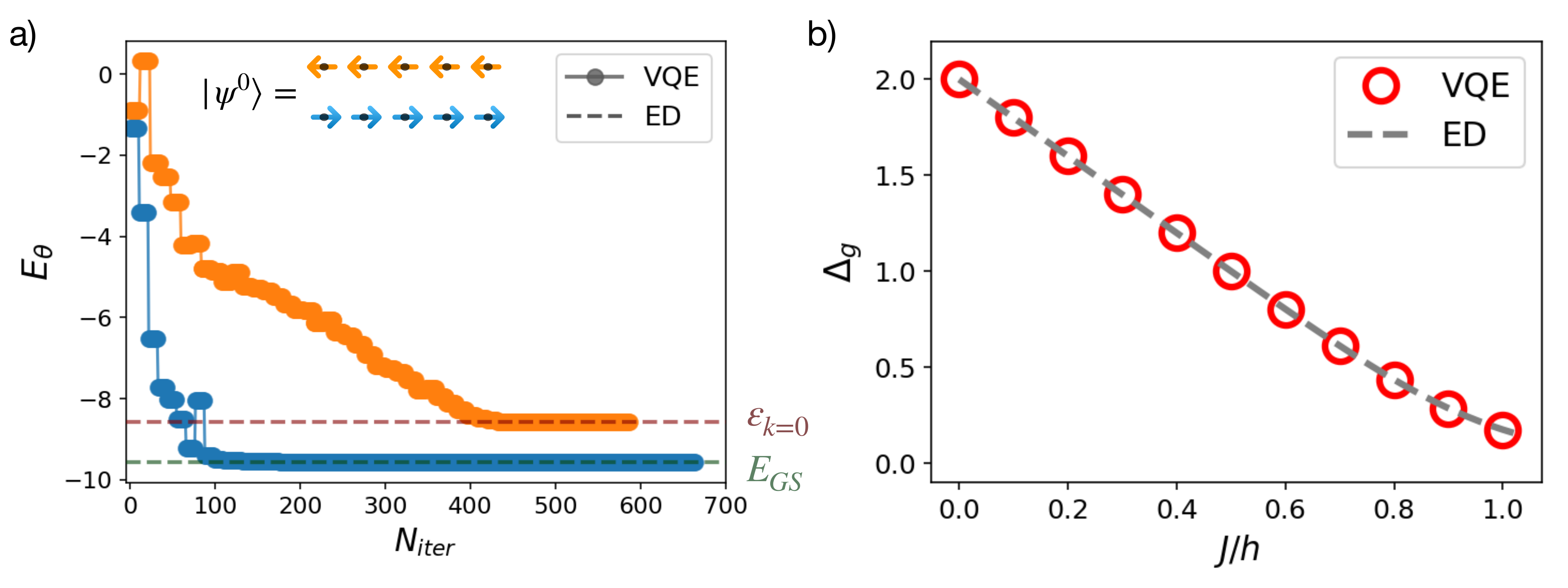}
    \caption{\justifying{\textbf{Extracting band gap using symmetry-guided VQE.}
    (a) convergence of energies $E_{\pmb{\theta}}$ as a function of iteration steps $N_{\rm iter}$ in VQE for initial states in different symmetry sectors, in the paramagnetic phase of the TFIM at $J/h = 0.5$. We used two translation invariant initial states with even (blue) and odd (orange) parities, specifically, the product states $\ket{\psi^0}=\otimes_{i=1}^N\ket{+}_i$ and $\ket{\psi^0}=\otimes_{i=1}^N\ket{-}_i$, respectively. Simulation in the even sector converges to the ground state with energy $E_{GS}$ (blue), whereas the odd sector yields the energy of the lowest magnon excitation $\varepsilon_{k=0}$. (b) Band gap $\Delta_g\equiv\varepsilon_{k=0}-E_{GS}$ as a function of $J/h$, evaluated by comparing VQE  runs for translation invariant initial states in the even and odd parity sectors (symbols). Comparison to exact diagonalization results (dashed line) shows excellent agreement.
    }}
    \label{fig:kspace_alg}
\end{figure*}

We note that for these initial states and with appropriately chosen parameters $\pmb{\theta}$, the circuit ~\eqref{eq:circuit} can realize the trotterization of a quantum quench, where the Hamiltonian is tuned from a parent Hamiltonian of the paramagnetic or ferromagnetic state, i.e., $H_{X}$ or  $H_{ZZ}$, to $H_{\rm TFIM}$, while preserving translation and parity symmetry. In particular,  increasing the circuit depth $d$ allows us to obtain better and better approximations of an adiabatic quench from either $H_{X}$ or  $H_{ZZ}$ to the Hamiltonian of interest $H_{\rm TFIM}$, except for the critical point $J=h$. This adiabaticity argument ensures that VQE can prepare the lowest energy state with respect to $H_{\rm TFIM}$ with the same symmetry properties as $\ket{\psi^0}$ to arbitrary precision, as long as the quantum circuit is sufficiently deep.

Before turning to the preparation of excited states, we briefly comment on constructing the approximate ground state with this circuit Ansatz, focusing on the limit of high magnetic field for simplicity. Here, the ground state belongs to the symmetry sector with even parity $p=+1$ and momentum $k=0$. A convenient initial state with these quantum numbers is the simple product state $\ket{\psi^0}=\otimes_{i=1}^N\ket{+}_i$, with $\ket{\pm}_i$ denoting the eigenstates of the operator $X_i$ throughout the paper, $X_i \ket{\pm}_i = \pm\ket{\pm}_i$. Note that $\ket{\psi^0}$ is the ground state of the parent Hamiltonian of the paramagnetic phase $H_{X}$.  As expected from the adiabaticity argument sketched above,  the VQE algorithm converges to the lowest energy eigenstate that matches the specified quantum numbers, i.e., to the ground state of $H_{\rm TFIM}$. The rapid convergence of the optimization process to the ground state energy $E_{GS}$ is illustrated in Fig. \ref{fig:kspace_alg}a.

We close this section by noting that while general momentum eigenstates do not have a product structure, making them an impractical choice for $\ket{\psi^0}$, the uniform state $k=0$ is an exception. This allows us to prepare the first excited state of the paramagnetic phase, corresponding to $k=0$ with energy $\varepsilon_{k=0}\equiv E_{p=-1,k=0}$, with a simplified VQE protocol.  This lowest excitation is characterized by the quantum numbers $k=0$ and an odd parity $p=-1$.  A convenient initial state within this symmetry sector is $\ket{\psi^0}=\otimes_{i=1}^N\ket{-}_i$, provided that the chain contains an odd number of sites. With this choice for $\ket{\psi^0}$, the VQE protocol indeed quickly converges to the first excited state as can be seen in Fig. \ref{fig:kspace_alg}a. Comparing the energies obtained for $\ket{\psi^0}=\otimes_{i=1}^N\ket{\pm}_i$ also allows us to extract the band gap between the ground state and the first excited band, since the minimal energy of the first band corresponds precisely to $k=0$. This band gap $\Delta_g$ is shown in  Fig. \ref{fig:kspace_alg}b, for different ratios $J/h$ within the paramagnetic phase.

We now turn to using VQE to prepare quasiparticles, gaining information on the full band, for arbitrary wave number $k$. We first discuss the case of magnon quasiparticles in the paramagnetic phase in Sec.~\ref{sec:magnon}, then explore the soliton quasiparticles of the ferromagnetic phase in Sec.~\ref{sec:dw}.

\section{VQE preparation of Magnon quasiparticles}\label{sec:magnon}

In this section, we illustrate how symmetry-guided VQE can be used to prepare the magnon quasiparticles of the Transverse Field Ising Model (TFIM) in the high magnetic field regime, $J/h < 1$ all the way up to the critical point $J/h = 1$.  Our discussion demonstrates the advantages of studying localized excitations on a quantum processor. These localized quasiparticle states encapsulate information about all momentum eigenstates forming the target excited band, opening the way to exploit quantum parallelism. We also comment on the efficiency of the VQE method, and explore various properties of the excited band and magnon quasiparticles that can be uncovered from these simulations.

\subsection{Symmetry guided VQE in the high field limit}

In this section we give more details on how to apply the general VQE framework of Sec.~\ref{subsec:vqe} to construct the quasiparticles in the paramagnetic phase of the TFIM, where the low lying excitations are magnons.  To develop an intuition, and to find a convenient choice for the initial state $\ket{\psi^0}$, it is instructive to first examine the exactly solvable limit $J=0$. As mentioned above, here the ground state is $\ket{\psi^{J=0}_{\rm GS}} = \otimes_{i=1}^N\ket{+}_i$, belonging to the symmetry sector of even parity, $p=1$. The lowest excited states are localized spin flips,
$$\ket{\psi^{J=0}_{-1,x}}=\ket{-}_x\otimes_{i\neq x}\ket{+}_i,$$ 
with odd parity $p=-1$.  All momentum eigenstates in the first excited band can be constructed from these degenerate spin flip excitations corresponding to different positions $x$ as 
$$\ket{\psi^{J=0}_{-1,k}} = \frac{1}{\sqrt{N}} \sum_{x=1}^N e^{-i k x} \ket{\psi^{J=0}_{-1,x}},$$
giving rise to a flat band of magnons. Here we consider periodic boundary conditions, so that $k=2\pi n/N$ with $n=0,...,N-1$.

For finite coupling $J$, but still within the paramagnetic phase, the Ising term $H_{ZZ}$ introduces a hopping term for localized magnon excitations, such that they cease to be eigenstates of the Hamiltonian. Consequently, all low energy eigenstates become extended, and the  band acquires dispersion. Importantly, the momentum eigenstates forming this band, $\ket{\psi_{-1,k}}$, are adiabatically connected to their $J=0$ counterpart $\ket{\psi^{J=0}_{-1,k}}$, retaining the same quantum numbers $k$ and $p=-1$. While the Hamiltonian does not support localized eigenstates anymore, it is still instructive to construct localized "Wannier states" of quasiparticles from the eigenstates $\ket{\psi_{-1,k}}$, with the  Wannier state localized around site $x$ given by 
\begin{equation*}
   \ket{\psi_{-1,x}} \equiv \frac{1}{\sqrt{N}} \sum_{k} e^{i k x} \ket{\psi_{-1,k}}.
\end{equation*}
The state $\ket{\psi_{-1,x}}$, despite not being an eigenstate of the Hamiltonian, can still be interpreted as a localized excitation, and carries all information on the full band. Specifically, each Wannier state $\ket{\psi_{-1,x}}$ can be understood as the renormalized counterpart of the spin flip excitation $\ket{\psi^{J=0}_{-1,x}}$, dressed by the Ising interactions, while retaining its odd parity, and the property that it has equal weight in all momentum sectors $k$.

We can now apply the VQE algorithm to construct the localized quasiparticle states $\ket{\psi_{-1,x}}$ of the lowest excited band. We choose the initial state as an exact spin flip excitation of the limit $J=0$,
\begin{equation}\label{eq:initial}
 \ket{\psi^0_x}\equiv\ket{\psi^{J=0}_{-1,x}}.
\end{equation}
We note that $\ket{\psi^0_x}$  is a product state that is easy to prepare in a quantum device, which is a crucial property for practical applications. The parametrized circuit $U(\pmb\theta)$ has the form~\eqref{eq:circuit}, resulting in the variational wave function
\begin{equation*}
    \ket{\psi_x(\pmb\theta)}\equiv U(\pmb{\theta}) \ket{\psi^0_x},
\end{equation*} 
and the cost function is given by the expectation value of the Hamiltonian  $H_{\rm TFIM}$, in accordance with Eq.~\eqref{eq:cost}. As we shall see below, with these choices, the adiabatic connection to the limit $J=0$ ensures the convergence of VQE to the desired magnon quasiparticle states for a sufficiently deep circuit. Indeed, in terms of the Fourier components of the localized variational state,
\begin{equation}
    \ket{\psi_k(\pmb \theta)} \equiv \frac{1}{\sqrt{N}} \sum_x e^{-i k x} \ket{\psi_x(\pmb \theta)},
\end{equation}
the cost function $E_{\pmb\theta}$ can be written as
\begin{equation*}
\begin{split}
    E_{\pmb\theta}&\equiv \bra{\psi_x(\pmb \theta)} H_{\rm TFIM} \ket{\psi_x(\pmb \theta)}\\
    &= \frac{1}{N} \sum_{k} \bra{\psi_{k}(\pmb \theta)} H_{\rm TFIM} \ket{\psi_k(\pmb \theta)} \\
    &\geq \frac{1}{N} \sum_{k} \bra{\psi_{-1,k}} H_{\rm TFIM} \ket{\psi_{-1,k}}.
\end{split}
\end{equation*}
Here we first used the translation invariance of the TFIM and the orthogonality of states with different momenta, $\bra{\psi_k(\pmb\theta)}\ket{\psi_{k^\prime}(\pmb\theta)}=\delta_{k,k^\prime}$, then the fact that the exact eigenstate $\ket{\psi_{-1,k}}$ gives the lowest energy in a given momentum sector $k$ and with odd parity $p=-1$. The lower bound can  be realized if it is possible to find a parameter set $\pmb{\theta}^*$ such that each momentum component  satisfies
\begin{equation}\label{eq:mincond}
    \ket{\psi_k(\pmb \theta^*)}=e^{i \phi_k}\ket{\psi_{-1,k}},
\end{equation}
with the factor $e^{i \phi_k}$ accounting for the phase freedom of eigenstates. In this case, the minimum $E_{\pmb\theta^*}$  is the average energy of  the excited band,
\begin{equation}\label{eq:Delta_av}
    \overline{\varepsilon}\equiv \dfrac{1}{N}\sum_k\varepsilon_k,
\end{equation}
where $\varepsilon_k$ is the magnon energy for wave number $k$.
With the circuit ansatz ~\eqref{eq:circuit}, the existence of $\pmb{\theta}^*$ satisfying ~\eqref{eq:mincond} is ensured  for deep enough circuits, by a reasoning analogous to the one used for the ground state in Sec.~\ref{subsec:vqe}. Namely, our choice of initial state ~\eqref{eq:initial} implies $\ket{\psi_k(\pmb \theta)}=U(\pmb{\theta})\ket{\psi^{J=0}_{-1,k}}$. Since adiabatic quenches between $H_X$ and $H_{\rm TFIM}$, respecting translation invariance and parity symmetry, transform $\ket{\psi^{J=0}_{-1,k}}$ into $\ket{\psi_{-1,k}}$ (up to a phase), and circuits of the form ~\eqref{eq:circuit} can approximate such quenches to arbitrary precision, the optimal solution ~\eqref{eq:mincond} indeed exists. Therefore, the VQE algorithm will converge towards a localized quasiparticle state,
\begin{equation}\label{eq:vqe_output}
    \ket{\psi_x(\pmb \theta^*)}=\frac{1}{\sqrt{N}} \sum_{k} e^{i k x+i \phi_k}\ket{\psi_{-1,k}}.
\end{equation}
 We note that the free phase factor $\phi_k$  appearing in this equation has important consequences for the output of the VQE. We will discuss this below in Sec.~\ref{sec:magnon_numerics}.

To summarize, the variational optimization ensures that the cost function is minimized at a parameter set $\pmb{\theta}^*$ such that each momentum component of $\ket{\psi_x(\pmb \theta^*)}$ is optimized to the correct magnon eigenstate within that momentum channel. This shows the effectiveness of the VQE in achieving quantum parallelism. By harnessing the power of unitary evolution, VQE can construct what are effectively `Wannier states' of magnon quasiparticles, making this approach to preparing and studying quasiparticle excitations particularly valuable. In the subsequent sections, we will discuss in detail the information that can be extracted from the variational Wannier states $\ket{\psi_x(\pmb \theta^*)}$.

\subsubsection*{Obtaining the full band dispersion}

As noted above, the localized quasiparticle state constructed via VQE contains information about all the momentum eigenstates, prompting us to consider how to extract this information. Below we demonstrate how to obtain the entire magnon dispersion from the prepared localized excitation $\ket{\psi_x(\pmb \theta^*)}$.

By using Eq.~\eqref{eq:vqe_output}, the magnon energy $\varepsilon_k\equiv \bra{\psi_{-1,k}}H_{\rm TFIM}\ket{\psi_{-1,k}}$ can be expressed in terms of localized quasiparticle states $\ket{\psi_x(\pmb \theta^*)}$ centered around different positions as follows,
\begin{equation}
\begin{split}
    \varepsilon_k&\equiv \bra{\psi_{-1,k}}H_{\rm TFIM}\ket{\psi_{-1,k}} \\
    &= \frac{1}{N} \sum_{x,x'} e^{ik(x'-x)} \bra{\psi_{x^\prime}(\pmb \theta^*)}H_{\rm TFIM}\ket{\psi_x(\pmb \theta^*)} \\
    &= \expval{H_{\rm TFIM}}{\psi_{x}(\pmb \theta^*)} \\
    &\quad\quad + \sum_{n=1}^{N-1} \cos (kn) \mel{\psi_{x+n}(\pmb \theta^*)}{H_{\rm TFIM}}{\psi_{x}(\pmb \theta^*)},
\end{split}
\label{eq:Ek_1}
\end{equation}
with arbitrary lattice site  $x$. Here, we used the translation invariance of the Hamiltonian to obtain the last equality. 

The formula ~\eqref{eq:Ek_1} allows us to extract the energy of each momentum channel from the output state of VQE. In practical implementations, the first term, $\expval{H_{\rm TFIM}}{\psi_{x}(\pmb \theta^*)}$, can be evaluated directly by performing measurements on the output state $\ket{\psi_x(\pmb \theta^*)}$. The matrix elements $f(n)\equiv \mel{\psi_{x+n}(\pmb \theta^*)}{H_{\rm TFIM}}{\psi_{x}(\pmb \theta^*)}$, however,  require a more involved experimental protocol. In principle, these terms can be reconstructed by performing full state tomography on the output state $\ket{\psi_x(\pmb \theta^*)}$. An alternative procedure involves first acting with the operators $H_X$ or $H_{ZZ}$ on $\ket{\psi_x(\pmb \theta^*)}$, and then applying the inverse of the variational circuit $U(\pmb{\theta}^*)$ on the resulting wave function, leading to the final state $U^{-1}(\pmb{\theta}^*)H_{X/ZZ}\ket{\psi_x(\pmb \theta^*)}$. Measuring this wave function in a spin-$z$ basis grants access to its overlap with different spin flip states $\ket{\psi^{J=0}_{-1,x+n}}$, which is precisely the information needed to reconstruct $f(n)$. While both of the protocols above are theoretically sound, they can be resource-intensive and pose practical challenges. We note that by observing that $f(n)$ can be rewritten as the expectation value of a unitary operator, one can also rely on another, potentially more efficient experimental procedure~\cite{CU_PhysRevX.7.021050}. This protocol circumvents the need for full state tomography at the expense of introducing an ancilla qubit, and involves the application of a controlled unitary gate between the system and the ancilla~\footnote{We note that the symmetry adapted variational protocol proposed in Ref. \cite{symm_adapt_PhysRevA.101.052340} involves a similar procedure, allowing in principle to imprint the desired momentum on the variational wave function prepared in the quantum device. A notable difference compared to our approach is the need to perform this operation in each step of the optimization process, In contrast, by working with Wannier states, we ensure that it has to be performed only once on the already optimized wave function to access the full dispersion relation.}.

Irrespective of the chosen measurement scheme, the practical implementation of the protocol required to extract the full dispersion relation remains challenging. In Sec.~\ref{sec:magnon_numerics} and  Sec.~\ref{sec:outlook}, we will discuss ways to partially circumvent these difficulties, by designing simple, practical protocols directly accessing important properties of the band, such as the average band gap and the band width. Turning to the numerical simulations presented in our paper, the computation of $f(n)$ is significantly more straightforward there. We can directly evaluate the overlap of localized quasiparticle states centered at different sites by translating the quasiparticle through a simple relabeling of the wavefunction indices.

The above discussion shows that a single VQE calculation, preparing a single Wannier state of quasiparticles, encodes information in an accessible way for an entire band in one run. We further demonstrate the potential of VQE, and discuss possible limitations and difficulties, by turning to numerical simulations below.

\subsection{Numerical simulations of Magnon quasiparticles}\label{sec:magnon_numerics}

In this section, we present our numerical simulations for the VQE approach introduced above to verify and elucidate different aspects of the algorithm. We explore how the magnon quasiparticles evolve within the paramagnetic phase, from small ratios $J/h$ towards the critical point $J/h=1$. We focus on 1D chains with an odd number of lattice sites, positioning the initial spin flip at the center, i.e., we choose the initial state according to ~\eqref{eq:initial}  with $x = (N+1)/2$. Our results show that the algorithm can accurately recover the Wannier states of quasiparticles, even when they are highly renormalized, right up to the critical regime. 

\subsubsection{Convergence Analysis}
\begin{figure}[t!]
    \centering
    \includegraphics[width=\linewidth]{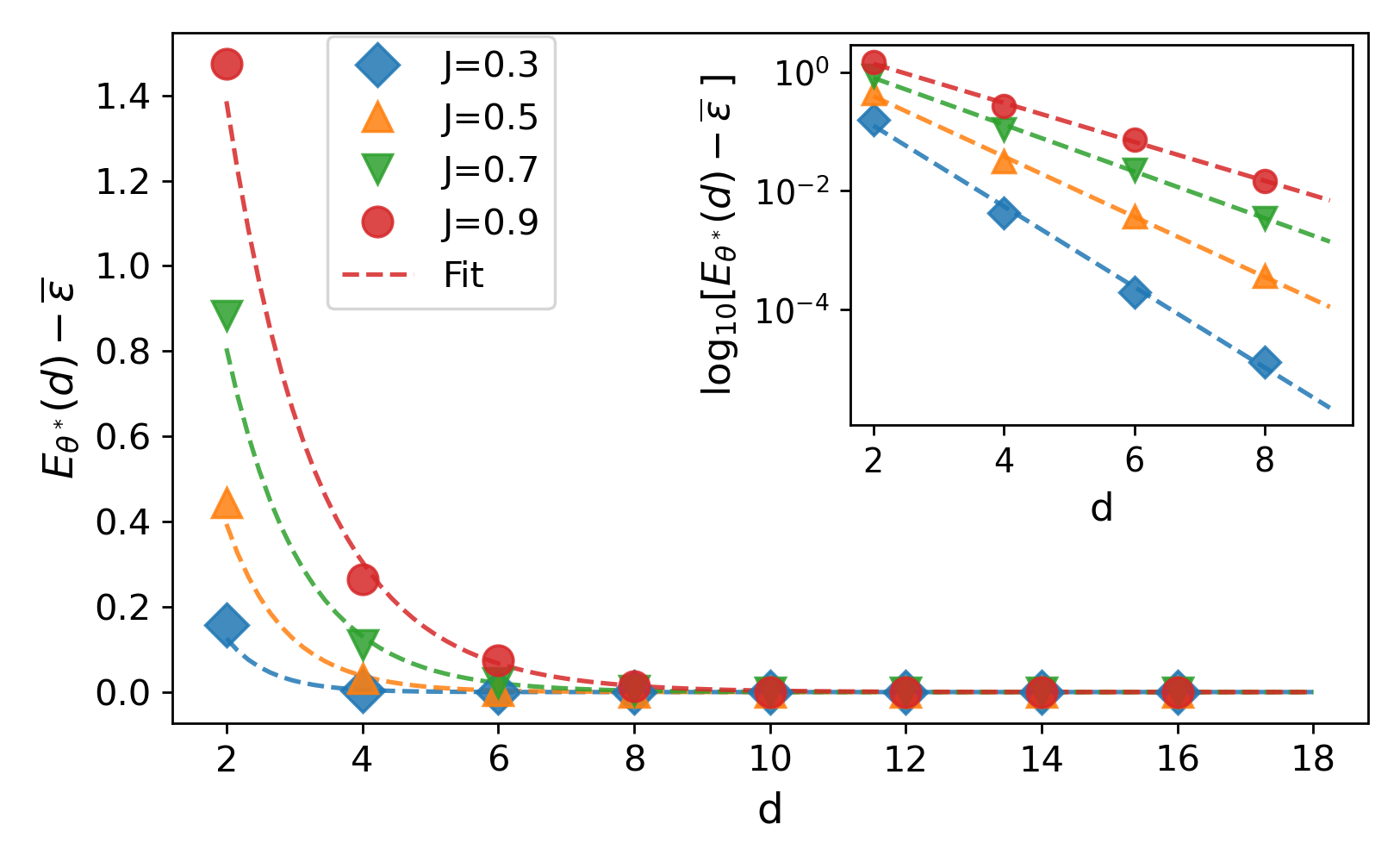}
    \caption{\justifying{\textbf{Precision of VQE vs. circuit depth.}
    Difference between the variational energy $E_{\pmb{\theta}^*}(d)$ of the localized magnon quasiparticle obtained in VQE, and the exact result $\overline{\varepsilon}$ evaluated via exact diagonalization, plotted as a function of circuit depth $d$ (symbols). Data shown for different ratios of $J/h$ in the paramagnetic regime. Exponential fits (dashed lines) demonstrate the exponential convergence of $E_{\pmb{\theta}^*}(d)$ towards $\overline{\varepsilon}$ with increasing circuit depth $d$. Inset shows the same data using logarithmic scale on the vertical axis, with a good linear fit confirming exponential dependence, ensuring excellent precision in VQE already at low depths for moderate system sizes. Here, we used system size $N=9$.
    }}
    \label{fig:convergence}
\end{figure}

We first examine the convergence properties of the VQE algorithm. Figure \ref{fig:convergence} shows  the converged energy $E_{\pmb{\theta}^*}$ for various depths of the unitary circuit $d$, compared to the average energy of the first excited band $\overline{\varepsilon}$, Eq.~\eqref{eq:Delta_av}. As expected from the adiabaticity arguments presented above, $E_{\pmb{\theta}^*}$ converges quickly towards the average magnon energy $\overline{\varepsilon}$ as $d$ increases.  Specifically, we find that the variational energy $E_{\pmb{\theta}^*}$  approaches the exact target energy $\overline{\varepsilon}$ exponentially as a function of  $d$, across the whole paramagnetic regime.  The observed exponential convergence suggests that for fixed moderate system sizes even relatively shallow circuits are well suited for variationally constructing the desired localized states of quasiparticles. We note, however, that the circuit depth required to obtain \textit{perfect} Wannier states is expected to approach a linear scaling with the system size $N$ in the vicinity of the critical point, based on the VQE results for the ground state of the TFIM.

\subsubsection{Estimating Magnon Energies}

\begin{figure}[t]
    \centering
    \includegraphics[width=\linewidth]{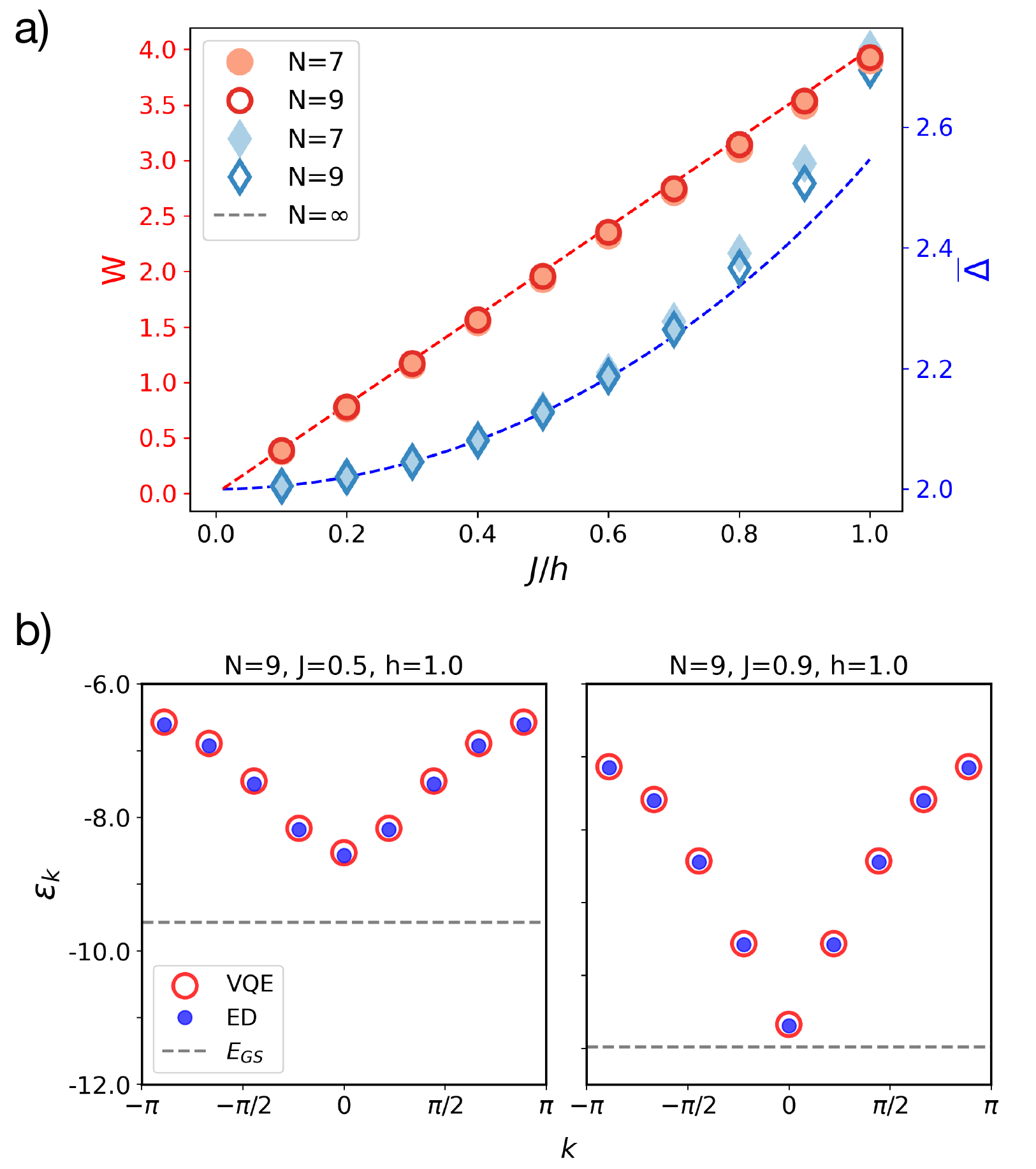}
    \caption{\justifying{
    \textbf{Magnon spectrum from VQE.} (a) Average band gap $\overline{\Delta}$ (blue symbols) and the width of the magnon band $W$ (red symbols) in the TFIM extracted from VQE, compared to exact results for the thermodynamic limit $N\to \infty$ (dashed lines), shown as a function of $J/h$ across the paramagnetic phase. The VQE results closely mirror the theoretical predictions of the thermodynamic limit, despite the small system sizes used here.
    (b) Full magnon band dispersion, extracted from the localized magnon quasiparticle state prepared via VQE using Eq.~\eqref{eq:Ek_1}. The VQE results (red symbols) are shown for two different ratios $J/h$, yielding a perfect match with exact diagonalization predictions (blue symbols). Ground state energy is also displayed for reference (dashed line). Here we used $N=9$.
    }}
    \label{fig:xbasis_vqe}
\end{figure}

Based on the convergence data depicted in Figure \ref{fig:convergence}, an appropriate circuit depth can be selected depending on the ratio $J/h$ and the precision required. As observed above, the quasiparticle energy $E_{\pmb{\theta}^*}$ represents the average energy across the `magnon' band. Thus, by subtracting the ground state energy from the energy of the localized quasiparticle prepared via VQE, we obtain an accurate estimate for the average band gap $\overline{\Delta}$ between the ground state and first magnon band. In the thermodynamic limit, this average band gap is given by
\begin{equation}\label{eq:Deltaav}
\overline{\Delta}^\infty = \int_0^{\pi} 2 \sqrt{h^2 + J^2 - 2 J h \cos{k}} \, dk.
\end{equation}
In  Fig. \ref{fig:xbasis_vqe}a, we show the VQE estimates for $\overline{\Delta}$ (blue symbols), compared to the exact expression in the thermodynamic limit ~\eqref{eq:Deltaav} (blue dashed line). Since the excitation energies obtained with VQE are practically exact for the small system sizes $N$ considered here, the deviations between the VQE data and Eq.~\eqref{eq:Deltaav} stem purely from finite size effects. Notably, even with these relatively small system sizes, the VQE results begin to align closely with those of the thermodynamic limit, indicating effective simulation capabilities. We also note that extracting $\overline{\Delta}$ is straightforward in practical applications, in contrast to the more refined procedure required to obtain the indiviual magnon energies $\varepsilon_k$.

As outlined above, the energies of full magnon band $\varepsilon_k$ can also be determined from the variational quasiparticle state $\ket{\psi_x(\pmb \theta^*)}$, albeit with a more complicated protocol. Figure \ref{fig:xbasis_vqe}b displays our numerical results for the magnon band dispersion, derived from the VQE algorithm using Eq.~\eqref{eq:Ek_1}, benchmarked against exact diagonalization for two different cases of $J/h$. In Figure\ref{fig:xbasis_vqe}a,  we compare the width of the magnon band, $W$, extracted from the full-band VQE dispersion (red symbols), to the theoretical bandwidth in the thermodynamic limit ($4J/h$, dashed line). Both plots demonstrate excellent agreement, highlighting the accuracy of the VQE approach even for the small system sizes considered.

While the extraction of the full band dispersion showcases the capability of VQE to capture detailed quasiparticle properties, it can be resource-intensive and challenging to implement in practice, as discussed earlier in Section~\ref{sec:magnon}A. To address this, we have identified a simpler and more practical protocol for extracting the magnon band width, $W$, directly from VQE, which we will outline in detail later in Section~\ref{sec:outlook}. This is achieved by considering initial product states in the spin-$x$ basis similar to Eq.~\eqref{eq:initial}, but with the location of the spin flip being in the coherent superposition of two possible positions $x_1$ and $x_2$. These examples demonstrate the versatility of VQE, allowing us to extract various properties of the magnon band with relatively simple procedures by modifying the initial state used in the algorithm.

\begin{figure*}[t!]
    \centering
    \includegraphics[width=\textwidth]{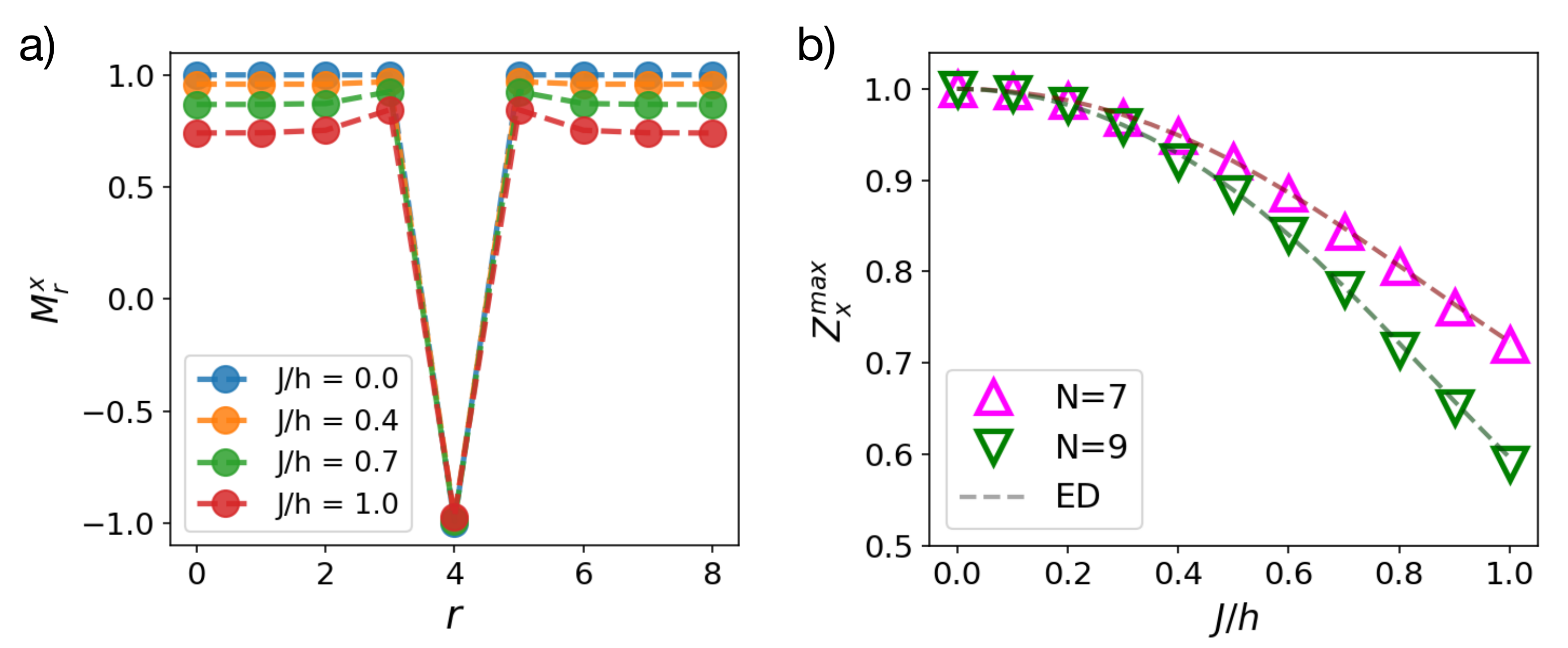}
    \caption{\justifying{\textbf{Wannier states prepared using VQE.}
    (a) Spatial magnetization profiles $M_r^x\equiv\langle X_r\rangle$ for the maximally localized magnon quasiparticle states, or Wannier states, prepared using VQE, shown for various ratios $J/h$ within the paramagnetic phase. The bare spin flip excitation at $J/h=0$ gets renormalized by interactions as $J/h$ is increased towards the phase boundary. Results are shown for $N=9$. Dotted lines are provided as a guide to the eye. (b) Quasiparticle weights of the maximally localized Wannier states obtained in VQE via post-selection (symbols), compared to exact predictions calculated through exact diagonalization (dashed lines), showing excellent agreement.
    }}
    \label{fig:xbasis-wavepackets}
\end{figure*}

\subsubsection{Width of localized Wannier states}
\label{sec:widths}

In this section, we explore the properties of Wannier states of magnon quasiparticles derived from VQE. The width of the maximally localized wave function that can be constructed from the extended momentum eigenstates forming the magnon band captures the broadening of the zero-width spin flip excitations of the limit $J=0$ due to Ising interactions. Thereby, this width encodes valuable information on interaction effects leading to the renormalization of magnons. While VQE is well suited for preparing localized Wannier states, the width of these states has no direct physical meaning due to the arbitrary phase factors $\phi_k$ appearing in Eq.~\eqref{eq:vqe_output}, resulting in different widths. Ideally, we aim to find a variational wave function with phases $\phi_k$ such that the state is maximally localized, thereby carrying direct information on the broadening caused by interactions. However, these phase factors are difficult to control in VQE. Below we discuss the behavior of phase factors that we observe in our numerical simulations in more detail and comment on its consequences.

A useful measure of the renormalization of  magnon states $\ket{\psi_{-1,k}}$ compared to the non-interacting limit $J=0$ is  the quasiparticle weight, defined as the overlap between $\ket{\psi_{-1,k}}$ and $\ket{\psi^{J=0}_{-1,k}}$,
\begin{equation}
    Z_k = \left| \bra{\psi^{J=0}_{-1,k}}\ket{\psi_{-1,k}} \right|^2.
\end{equation}
At $J=0$, $Z_k=1$ for all momenta; however, as we approach the critical point $J/h \to 1$, the quasiparticle becomes increasingly renormalized leading to a decrease in $Z_k$.

The quasiparticle weight for a localized Wannier state $\ket{\psi_x(\pmb \theta^*)}$ constructed via VQE, Eq.~\eqref{eq:vqe_output},  can be defined analogously,   as the overlap with the single spin flip excitation of the  limit $J=0$ centered around the same lattice site $x$,
\begin{equation}
\begin{split}
    Z_x &= \left| \braket{\psi_{-1,x}^{J=0}}{\psi_{x}(\pmb \theta^*)} \right|^2 = \left| \frac{1}{N} \sum_{k,k'} \braket{\psi_{-1,k}^{J=0}}{\psi_{k'}(\pmb \theta^*)} \right|^2 \\
    &= \frac{1}{N^2} \left| \sum_k e^{i \phi_k} \braket{\psi_{-1,k}^{J=0}}{\psi_{-1,k}} \right|^2\\
    &\leq  \left( \frac{1}{N}\sum_k \left| \braket{\psi_{-1,k}^{J=0}}{\psi_{-1,k}}\right| \right)^2\\
    &= \left( \frac{1}{N}\sum_k \sqrt{Z_k}\right)^2\equiv Z_x^{\rm max}.
\end{split}
\end{equation}
As noted above, the upper limit $Z_x^{\rm max}$ can be interpreted as a measure of interaction effects, dressing the bare spin exictations of the limit $J=0$. Indeed, a larger $Z_x$ indicates a greater overlap with the bare single spin flip state and, therefore, a more localized quasiparticle. To describe a maximally localized Wannier state of the magnon band, the phases $\phi_k$ in the variational wave function have to align in such a way that the terms  $e^{i \phi_k} \braket{\psi_{-1,k}^{J=0}}{\psi_{-1,k}}$ all have the same phase, across every momentum channel. According to the expression above, the weight $Z_x^{\rm max}$ corresponding to this maximally localized wave function is given by a generalized mean of the quasiparticle weights $Z_k$ over all momentum channels $k$.

In the VQE algorithm for localized excitations discussed in Sec.~\ref{subsec:vqe}, each momentum state is optimized only up to a phase, with nothing in the circuit design inherently preventing each $k$-channel from acquiring a random phase. Consequently, as the different $k$-channels can accumulate various phases, the Wannier state prepared with VQE may spread out, diminishing the quasiparticle weight $Z_x$. We explored this effect by examining the statistics of relative phases acquired in independent VQE runs. Our results are presented in Appendix~\ref{app:phase}, and show that in practice, for the small system sizes considered here, the distribution of relative phases is heavily skewed towards zero, implying a distribution of $Z_x$ with high weight close to its maximal value $Z_x^{\rm max}$. We note that the distribution becomes broader in the vicinity of the critical point $J=h$. We attribute the sharp distribution of relative phases, centered around zero, to our choice of initial state ~\eqref{eq:initial}, corresponding to a perfectly localized spin flip excitation with the phases of different momentum components aligned. This favors small relative phases in the output state $\ket{\psi_x(\pmb \theta^*)}$, especially deep in the paramagnetic phase where the circuit required to prepare the Wannier state is shallow. 

By relying on the observations above, for the small system sizes considered in this paper, we can employ the strategy of post-selection to find $Z_x^{\rm max}$, by selecting the variational wavefunction with the highest quasiparticle weigth from a small number of independent VQE runs. Figure~\ref{fig:xbasis-wavepackets}a shows the maximally localized Wannier states of the magnon quasiparticles obtained via this post-selection, for different interaction strengths $J$, with different levels of renormalization compared to bare spin flip excitations of the limit $J=0$. Specifically, increasing $J/h$ changes the uniform background magnetization set by the ground state, while also leading to the broadening of these localized quasiparticle states.
The quasiparticle weights  $Z_x^{\rm max}$ corresponding to such maximally localized wave functions are displayed in Fig.~\ref{fig:xbasis-wavepackets}b as a function of $J/h$ across the paramagnetic regime. We find a good agreement between the maximum obtained from VQE simulation (symbols) and the exact result calculated with exact diagonalization (dashed lines), demonstrating the efficiency of post-selection. We note, however, that a more careful analysis of the system size dependence of the phase distribution will be required for practical applications targeting moderate to large system sizes.  

\section{VQE preparation of Soliton quasiparticles}\label{sec:dw}

We now turn to the ferromagnetic phase of the TFIM, and discuss how to apply VQE to prepare soliton quasiparticles. In order to study single soliton excitations, we need to slightly modify the procedure presented in Sec.~\ref{subsec:vqe}, since a periodic chain with the TFIM Hamiltonian ~\eqref{eq:ham} only supports \textit{pairs} of domain wall excitations. The conventional solution is to change the periodic boundary conditions to open boundaries; the drawback of this approach is that the open boundaries break translation invariance, leading to strong finite size effects, particularly for the small system sizes considered here. This motivated us to present an alternative construction realizing twisted boundaries. Importantly, the twisted TFIM introduced in this approach supports single soliton states, while it retains translation symmetry in the form of a modified translation operator that still commutes with the Hamiltonian. Therefore the single soliton states obtained in this setting show much weaker finite size effects than their counterparts realized with open boundaries.

\begin{figure*}
    \centering
    \includegraphics[width=\linewidth]{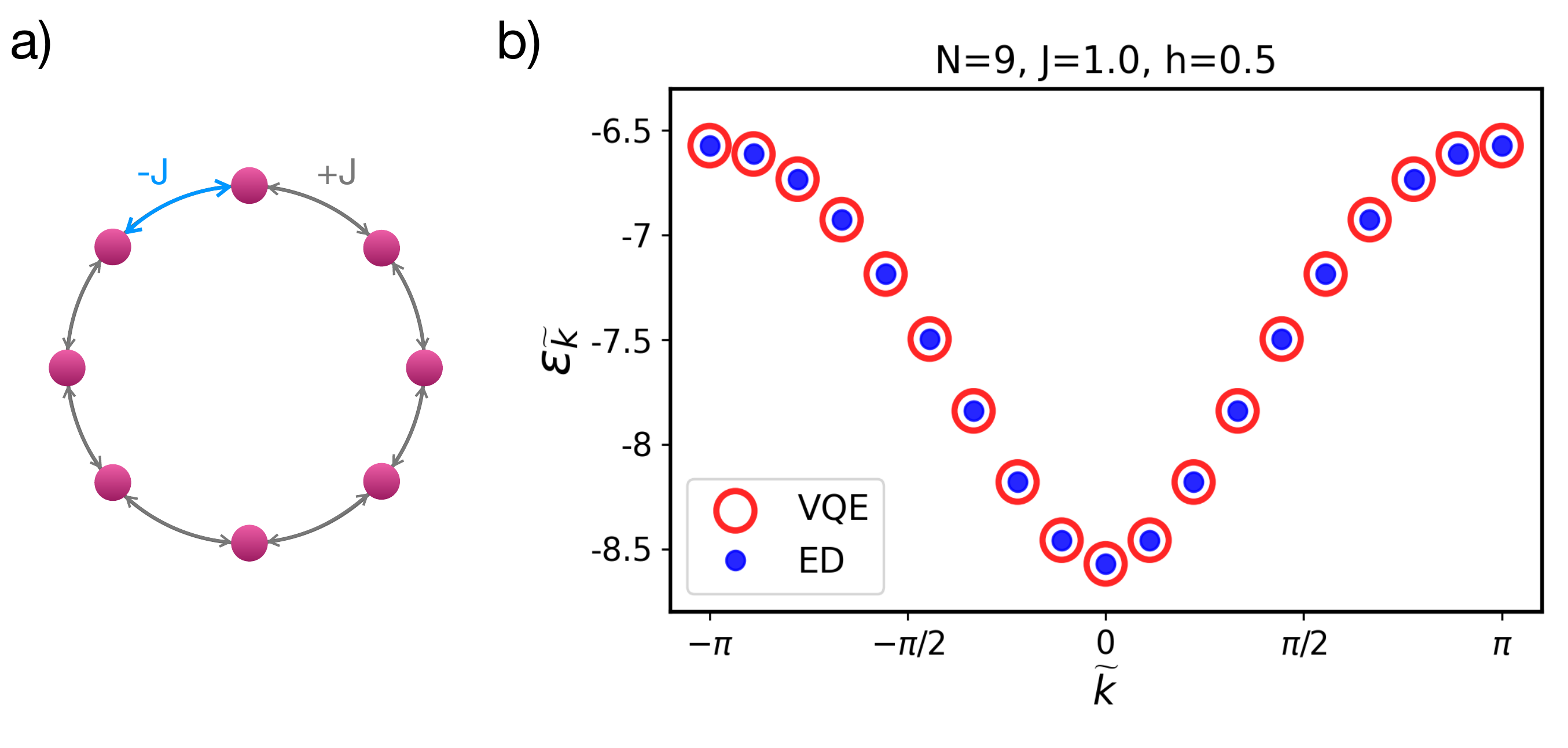}
    \caption{\justifying{\textbf{Symmetry guided VQE for soliton quasiparticles.}
    (a) Twisted TFIM, with the sign of $J$ flipped on a single bond, $J_N=J\rightarrow J_N=-J$, supporting single domain wall excitations in the system and allowing us to study single soliton quasiparticles with VQE. (b) Full soliton dispersion obtained from the Wannier states prepared with VQE via Eq.~\eqref{eq:Ek_1} (red symbols), compared to exact diagonalization performed on the twisted model ~\eqref{eq:twist} (blue symbols), showing excellent agreement. We used $J=1$ and $h=0.5$, with system size $N=9$.
    }}
    \label{fig:disp_zbasis}
\end{figure*}

Specifically, we isolate a single soliton excitation by considering a twisted version of the TFIM on a periodic chain, with one of the bonds flipped from ferromagnetic to antiferromagnetic (see Fig.~\ref{fig:disp_zbasis}a),
\begin{equation}\label{eq:twist}
    H_{\rm tTFIM} = - \sum_{i=1}^{N} J_i Z_i Z_{i+1} - h \sum_{i=1}^{N} X_i,
\end{equation}
where $J_i = J$ for $i < N$ and $J_N = -J$. 
As we discuss in more detail below, the low energy manifold of $H_{\rm tTFIM}$ is spanned by configurations with a single soliton, precisely the quasiparticles we aim to study. For sufficiently large system sizes $N$, finite size effects become negligible and the single soliton Wannier states of the twisted model faithfully represent the localized elementary excitations of the TFIM in the thermodynamic limit.

Importantly, while this choice of coupling in the twisted model breaks the translation invariance of the original TFIM ~\eqref{eq:ham}, we can define a modified translation operator, $\tilde{T} = T X_N$, that still commutes with the Hamiltonian.
As detailed below, the presence of this generalized translation symmetry can be exploited in a way analogous to the role of conventional translation invariance in Sec.~\ref{subsec:vqe}, to target the low energy manifold hosting single soliton states. The spectrum of the twisted TFIM is detailed in Appendix~\ref{app:spectrum}. As before, we will focus on a chain with an odd number of sites.

The modified translation operator satisfies $\tilde{T}^N = P = \prod_i X_i$, implying $\tilde{T}^{2N} = 1$. Therefore, the eigenvalues of $\tilde{T}$ are given by $e^{i\tilde{k}}$, labelled by the generalized wave numbers  $\tilde{k} = \pi m/N$ with $m \in \mathbb{Z}$. The eigenstate labelled by $\tilde{k}= \pi m/N$ has a well defined defined parity due to $\tilde{T}^N = P$, namely $p=(-1)^m$.

Similarly to the strategy followed in Sec.~\ref{sec:magnon} to construct magnon quasiparticles, it is instructive to first examine the simple solvable limit $h=0$, and use it as guidance for choosing a convenient initial state $\ket{\psi^0}$ for VQE. As we shall see, at $h=0$, there is a correspondence between the \textit{ground state} manifold of the twisted model and the \textit{quasiparticle excitations} we aim to study. Specifically, the ground state manifold is spanned by $2N$ degenerate ground states, and  we can choose basis states that correspond to perfectly localized, bare domain wall excitations as follows. We start from a state polarized in the spin-$z$ direction,
\begin{equation}\label{eq:dw_loc0}
  \ket{\psi^{ h=0}_{\sigma=+,x=0}} = \otimes_{i=1}^N\ket{\uparrow}_i,   
\end{equation}
with $Z_i\ket{\uparrow}_i=\ket{\uparrow}_i$. Here we introduced the subscript $\sigma=\pm$, referring to the up / down orientation of the rightmost spin at site $N$. The wave function~\eqref{eq:dw_loc0} is one of the ground states of the twisted Hamiltonian ~\eqref{eq:twist}, and hosts a frustrated bond at $x=0$,  between sites $N$  and 1, thereby representing a domain wall localized at $x=0$.   We can generate  $2N$  states describing localized domain walls from state~\eqref{eq:dw_loc0} via the generalized translation $\tilde{T}$ as
\begin{equation}\label{eq:dw_loc}
    \ket{\psi^{h=0}_{\sigma_n,x_n}} = \tilde{T}^n \ket{\psi^{h=0}_{\sigma=+,x=0}},
\end{equation}
with $\sigma_n=+$ for $n<N$, $\sigma_n=-$ for $n\geq N$, and $x_n=n$ mod $N$.  This way we obtain a complete basis for the ground state manifold of ~\eqref{eq:twist}, with all states corresponding to bare domain wall quasiparticles localized at bond $x_n=1,...,N$ in the original untwisted model. We note that the states ~\eqref{eq:dw_loc} do not have a well defined parity, in contrast to the localized spin flip excitations considered in Sec.~\ref{sec:magnon}, at the simple solvable point of the paramagnetic phase $J=0$. One can construct localized domain wall configurations with parity symmetry from the combinations $\ket{\psi^{h=0}_{+,x}}\pm\ket{\psi^{h=0}_{-,x}}$, however, these states are not practical initial states for VQE, since it is not easy to prepare them due to their GHZ structure. For this reason, as we demonstrate below, it will be more convenient to work with the product states~\eqref{eq:dw_loc}. We can construct momentum eigenstates from the wave functions~\eqref{eq:dw_loc} as
\begin{equation}\label{eq:dw_k}
    \ket{\psi^{h=0}_{\tilde k}} = \frac{1}{\sqrt{2 N}} \sum_{n=1}^{2N} e^{-i \tilde k n} \ket{\psi^{h=0}_{\sigma_n,x_n}},
\end{equation}
giving rise to a flat band similarly to the case of spin flip excitations in opposite simple solvable limit $J=0$. As noted above, these states have a well defined parities $p=e^{i\tilde{k}N}$. 

Away from the limit $h=0$, the transverse field generates a hopping term for domain walls. As a result, the band becomes dispersive, and the bare domain wall quasiparticles get renormalized, analogously to the broadening of spin flip excitations due to Ising interactions in the paramagnetic phase. Importantly, in perfect analogy with adiabaticity arguments applied for the magnon excitations of the paramagnetic phase, the emerging soliton quasiparticles with wave functions $\ket{\psi^{h}_{\tilde k}}$ remain adiabatically connected to the bare domain wall excitations of the limit $h=0$ across the whole ferromagnetic phase, i.e., adiabatic quenches from $h=0$ to the desired $h$ transform the wave functions into each other up to a phase, an observation crucial for the convergence of VQE presented below. Before turning to this VQE method, we can gain intuition about the nature of soliton quasiparticles via first order perturbation theory in $h$. For  $h \ll J$,  this predicts the soliton energies
$$\varepsilon_{\tilde k} = -(N-2)J - 2h \cos \tilde k + O(h^2),$$
with the ground state of the twisted model~\eqref{eq:twist} corresponding to the uniform state $\tilde{k}=0$ with even parity $p=1$. The only other non-degenerate eigenstate corresponds to  $\tilde k=\pi$ with $p=-1$ for the case of odd $N$ considered here, while all other eigenstates come in two-fold degenerate pairs $\pm\tilde{k}$.

With these insights we can now generalize the VQE approach presented in Sec.~\ref{sec:magnon} to the soliton quasiparticles of the ferromagnetic phase. We choose the initial state of the VQE algorithm as an easy-to-prepare product state $\ket{\psi^{ h=0}_{\sigma,x}}$ according to Eq.~\eqref{eq:dw_loc}, such that 
\begin{equation*}
 \ket{\psi_{\sigma,x}(\pmb\theta)}= U(\pmb{\theta}) \ket{\psi^{ h=0}_{\sigma,x}}.   
\end{equation*}
We use a cost function similar to Eq~\eqref{eq:cost}, but with the twisted Hamiltonian~\eqref{eq:twist},
\begin{equation*}
    E_{\pmb\theta}\equiv \bra{\psi_{\sigma,x}(\pmb \theta)} H_{\rm tTFIM} \ket{\psi_{\sigma,x}(\pmb \theta)}.
\end{equation*}
It is straightforward to repeat the arguments of Sec.~\ref{sec:magnon} to show that 
\begin{equation*}
 E_{\pmb\theta}\geq \dfrac{1}{2N}\sum_{\tilde{k}}\varepsilon_{\tilde{k}},
\end{equation*}
with the lower bound realized if it is possible find a parameter set $\pmb{\theta}^*$ such that
\begin{equation*}
     \ket{\psi_{\tilde k}(\pmb \theta^*)}=e^{i \phi_{\tilde k}}\ket{\psi^{h}_{\tilde k}},
\end{equation*}
for all momentum components 
\begin{equation*}
    \ket{\psi_{\tilde k}(\pmb \theta)} \equiv \frac{1}{\sqrt{2 N}} \sum_{n=1}^{2N} e^{-i \tilde{k} n} \ket{\psi_{\sigma_n,x_n}(\pmb \theta)}.
\end{equation*}
Therefore, VQE is well suited to prepare the momentum eigenstate $\ket{\psi^{h}_{\tilde k}}$ in each channel $\tilde{k}$, up to an arbitrary phase factor $\phi_{\tilde k}$.

The convergence of VQE to a localized Wannier state of soliton quasiparticles is again guaranteed by the adiabatic connection to the limit $h=0$. Similarly, the protocol presented in Sec.~\ref{sec:magnon} can be easily extended to show that the Wannier state $\ket{\psi_{\sigma,x}(\pmb\theta)}$ contains information on the whole soliton band, in particular, the full soliton dispersion is in principle accessible. We demonstrate this through numerical simulations, with the predictions of VQE for the soliton dispersion plotted in Fig.~\ref{fig:disp_zbasis}b, showing good agreement with exact diagonalization results obtained for the twisted model ~\eqref{eq:twist}. While a direct comparison to the spectrum of the untwisted TFIM ~\eqref{eq:ham} of the same system size is hindered by finite size effects for the small $N$ considered here, we find that the dispersion relation of the twisted model ~\eqref{eq:twist} is remarkably close to the exact analytical predictions for the thermodynamic limit, showing the advantages of this construction. We note that in perfect analogy with the case of magnon quasiparticles, the width of maximally localized Wannier states of solitons carries crucial information on the renormalization of bare domain wall excitations. Due to the phase freedom, accessing these maximally localized states in VQE is not straightforward, and requires post-selection similar to the one discussed in Sec.~\ref{sec:magnon_numerics}.

\section{Extensions and future scope}\label{sec:outlook}

In this paper, we have extended the variational quantum eigensolver formalism to capture low energy quasiparticle excitations. We have shown that by exploiting translation symmetry, and potential other abelian symmetries of the Hamiltonian, VQE can prepare localized Wannier states of quasiparticles, encoding all information on the full band and thereby realizing quantum parallelism. We have benchmarked this protocol via numerical simulations performed on the transverse field Ising model, hosting different types of quasiparticles in its paramagnetic and ferromagnetic phase. We have shown that VQE allows us to access the full dispersion of both magnon and domain wall quasiparticles, and encodes valuable information on how interactions renormalize the bare excitations of easily solvable non-interacting limits. These results provide important input for potential applications of near-term noisy intermediate scale quantum devices, and establish them as promising platforms for studying quasiparticles in strongly interacting quantum systems, thereby yielding direct predictions for essential experimentally measured quantities, such as transport properties. We conclude by briefly commenting on various potential extensions and future applications of this protocol. 

\subsection{Extension of VQE to different initial states}

\begin{figure}
    \centering
    \includegraphics[width=\columnwidth]{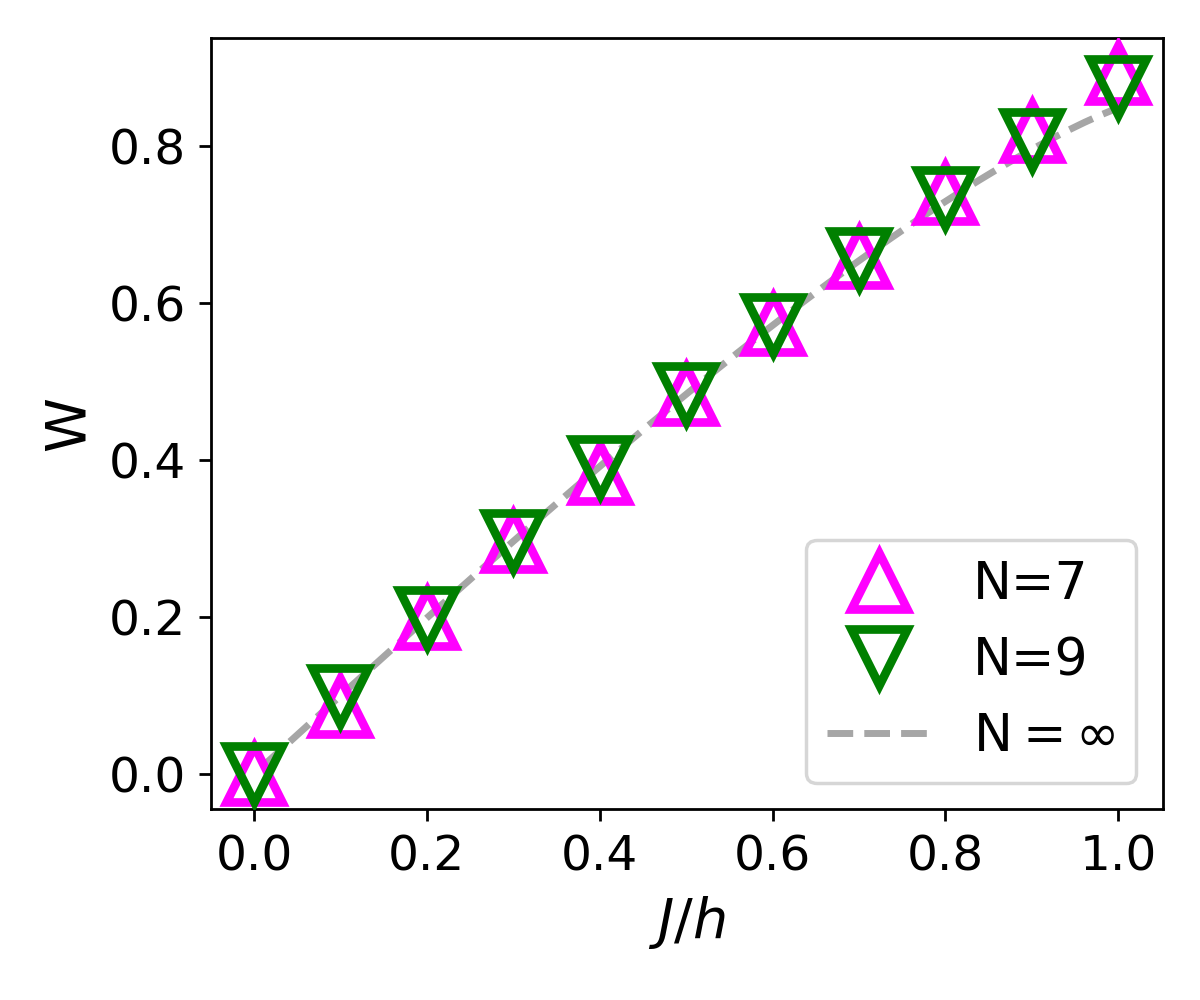}
    \caption{\justifying{\textbf{Efficient extraction of magnon band width using VQE:} Magnon band width $W$ in the TFIM extracted from VQE simulations using a superposition-based initial state, compared to exact results in the thermodynamic limit ($N \to \infty$, dashed line), plotted as a function of $J/h$ in the paramagnetic phase. The close agreement between the VQE results and the theoretical predictions demonstrates the accuracy of this approach, even for small system sizes, and its practicality for directly extracting bandwidth without requiring the full band dispersion.}}
    \label{fig:width}
\end{figure}

As discussed in previous sections, the Wannier states prepared in VQE carry information on the full quasiparticle band. However, while accessing the full band dispersion is possible in principle, implementing the suggested procedure might be challenging in practice. Therefore, proposing simpler protocols that can yield information on the band, without the need to extract the full band dispersion, are valuable for potential applications. Furthermore, the VQE initialized with a single bare spin flip or domain wall quasiparticle does not provide insight into the interaction between quasiparticles; protocols that grant access to higher-order processes involving multiple quasiparticles can reveal crucial information on strongly interacting many-body systems.

With these motivations, understanding the output of the VQE algorithm for different initial states, possibly involving multiple bare excitations, is an interesting future research direction. As a simpler example, here we revisit the paramagnetic phase of the TFIM,  and demonstrate that a slight change in the initial state used in VQE allows us to extract the width of the magnon band directly, without the need to reconstruct the full magnon dispersion. To this end,  we consider a VQE algorithm analogous to the one presented in Sec.~\ref{sec:magnon}, with the initial state now chosen as
\begin{equation}\label{eq:newinitial}
    \ket{\psi_{x_1,x_2}^0}=\dfrac{\ket{-}_{x_1}\otimes\ket{+}_{x_2}+\ket{+}_{x_1}\otimes\ket{-}_{x_2}}{\sqrt{2}}\otimes_{i\neq x,x+1}\ket{+}_i.
\end{equation}
Similarly to the initial state used in Sec.~\ref{sec:magnon}, $\ket{\psi_{x_1,x_2}^0}$ still encodes a single spin flip, but the wave function is a coherent superposition of a state where the spin flip is at site $x_1$, and one where it is at site $x_2$. Importantly, this initial state can still be easily prepared in practice, by entangling the spins at sites $x_1$ and $x_2$ into a Bell pair along the spin-$x$ axis.
We denote the output of the unitary circuit~\eqref{eq:circuit} by
\begin{equation*}
    \ket{\psi_{x_1,x_2}(\pmb \theta)}=U(\pmb \theta)\ket{\psi_{x_1,x_2}^0}.
\end{equation*}
With this choice of initial state, the cost function, taken to be expectation value of the Hamiltonian $H_{\rm TFIM}$ as before, satisfies,
\begin{equation}
\begin{split}
    &\expval{H_{\rm TFIM}}{\psi_{x_1,x_2}(\pmb \theta)} \\
    & \;\;=\frac{1}{2} \left[  \expval{H_{\rm TFIM}}{\psi_{x_1}(\pmb{\theta})} + \expval{H_{\rm TFIM}}{\psi_{x_2}(\pmb{\theta})} \right.\\
    &\qquad\qquad+ \left.\left(\braket{\psi_{x_2}(\pmb{\theta})}{H_{\rm TFIM}|\psi_{x_1}(\pmb{\theta})} + \text{h.c.}\right)\right] \\
    &  \;\;=\frac{1}{N} \left[\sum_k \expval{H_{\rm TFIM}}{\psi_{k}(\pmb{\theta})}\right. \\
    &\qquad +\left.\frac{1}{2} \sum_{k,k^\prime} \left( e^{ikx_1-ik'x_2} \braket{\psi_{k}(\pmb{\theta})}{H_{\rm TFIM}|\psi_{k^\prime}(\pmb{\theta})} + \text{h.c.} \right) \right]\\
    &\;\;= \frac{1}{N} \sum_k \left( 1 + \cos(ky)\right) \expval{H_{\rm TFIM}}{\psi_{k}(\pmb{\theta})},
\end{split}
\end{equation}
with $y=x_1-x_2$. Here $\psi_{x}(\pmb{\theta})$ and $\psi_{k}(\pmb{\theta})$ are the variational wave functions defined in Sec.~\ref{sec:magnon}, for an initial state with a bare spin flip excitation at fixed position $x$. Since $1+ \cos(ky)\geq 0$, the cost function is minimized when each term $\expval{H_{\rm TFIM}}{\psi_{k}(\pmb{\theta})}$ is minimized independently,
\begin{equation*}
\begin{split}
     &\expval{H_{\rm TFIM}}{\psi_{x_1,x_2}(\pmb \theta)}\\
     &\qquad\geq\frac{1}{N} \sum_k \left( 1 + \cos(ky)\right) \expval{H}{\psi_{-1,k}}\\
     &\qquad=\overline{\varepsilon} + \frac{1}{N} \sum_k \cos(ky) \expval{H}{\psi_{-1,k}}.
\end{split}
\end{equation*}
The first term here is the average energy of the magnon band. This energy is straightforward to extract from the original VQE algorithm discussed in Sec.~\ref{sec:magnon}, since it coincides with the minimum of the cost function in that formulation (see Fig.~\ref{fig:xbasis_vqe}).  Therefore, subtracting $\overline{\varepsilon}$ from the minimum of the cost function obtained for the new initial state~\eqref{eq:newinitial} grants us easily accessible additional information of the magnon band, without the need to implement a complicated protocol to extract the full dispersion. For example, choosing $y = 1$ allows us to estimate the width of the magnon band $W$, 
\begin{equation*}
\begin{split}
    W 
    &\equiv  -\frac{1}{N} \sum_k \cos(k) \expval{HH_{\rm TFIM}}{\psi_{-1,k}} \\
    &\to -\int_0^{\pi} 2 \sqrt{h^2 + J^2 - 2 J h \cos{k}} \, \cos{k}\, dk \;\; \text{for} \; N\to \infty.
\end{split}
\end{equation*}
Indeed, the expression in the second line is the exact result for the width of the magnon band in the thermodynamic limit. 

In Figure \ref{fig:width}, we show our numerical results for $W$, obtained from VQE according to the procedure outlined above. We find that VQE results (red symbols), calculated for small system sizes, match the exact expression in the thermodynamic limit with almost no finite size effects. This suggests an interesting cancellation of finite size effects in the difference of cost functions, corresponding to VQE with initial states ~\eqref{eq:initial} and ~\eqref{eq:newinitial}, respectively.

The simple example above demonstrates the potential in exploring VQE with different initial states, to design practical protocols that capture various properties of quantum many-body systems. Exploring these possibilities further is an exciting open research direction with imminent practical relevance.

\subsection{Future Scope}

In this work,  we have focused on applying VQE to prepare quasiparticle states in a one-dimensional spin system. While our results demonstrate the potential of this approach in preparing localized quasiparticle states, extending this framework to other many-body settings, potentially hosting more intricate excitations, is crucial for future applications. Specifically, a similar formalism should be well suited for addressing the excitations of higher, two and three dimensional systems. A particularly interesting question to explore is to study topologically non-trivial quasiparticles in these settings, similarly to how modifying the TFIM and replacing conventional translation symmetry with a generalized translation operator allowed us to capture soliton excitations in the ferromagnetic phase of the 1d transverse field Ising chain. Furthermore, here we have focused on quantum many-body systems consisting of bosonic particles.   Extending the method to capture fermionic quasiparticles in Hubbard models and other fermionic systems is an exciting future direction with utmost practical relevance.

As briefly noted in the previous subsection, besides exploring different types of quasiparticles, the VQE approach presented here can also be extended to address the interactions between quasiparticle excitations. As long as the eigenstates involving two quasiparticles, carrying information on these interactions, are either well separated in energy or distingushed by symmetry from the single particle band, they can in principle be targeted by VQE. Characterizing these interactions could provide valuable insight into the structure of a larger set of eigenstates in the low energy spectrum, admitting a description in terms of a few interacting quasiparticles. We note, however, that the present VQE approach is tailored to target the quasiparticles governing the low energy behavior, and going beyond  states containing a few quasiparticles to study the denser high energy spectrum becomes increasing more difficult.

Another essential question to explore in future works is the implementation of these protocols on currently available quantum simulator platforms, such as settings involving diamond NV centers~\cite{Childress2013,Lee_2017}, or superconducting circuits such as the quantum processors of Google~\cite{Arute2019}. For these applications, it is crucial to study the robustness of the algorithm against experimental noise and imperfections that are inevitably present in all devices. Applying active error correction might be necessary to obtain reliable results on these architectures, warranting the construction of efficient error-correcting protocols~\cite{Roffe2019}. Several recent works have addressed similar questions in the case of ground state preparation, but extending these results to the quasiparticle states considered here remains an interesting open problem.  Finally, exploring the role of measurements in the preparation of Wannier states of quasiparticles is another exciting future research direction. In the current VQE algorithm, measurements only enter as inputs for the classical optimizer updating the parameters of the unitary circuit. Recent works have started to explore the potential of mid-circuit measurements to generate long-range entanglement and topological states with shallow quantum circuits, allowing to prepare ground states that would require applying deep quantum circuits in a purely unitary setting~\cite{Briegel2001,Piroli2021,Verresen2024}. Extending these ideas for quasiparticle states is an open question of imminent practical relevance. We leave the exploration of these exciting research directions for future work.

\begin{acknowledgments}
This research was supported in part by grant NSF PHY-2309135 to the Kavli Institute for Theoretical Physics (KITP). I.L. acknowledges support from the Gordon and Betty Moore Foundation through Grant GBMF8690 to UCSB.  R.J. was supported by the UCSB NSF Quantum Foundry through Q-AMASE-i program award number DMR-1906325, supplemented by the NSF CMMT program under Grant No. DMR-2419871. L.B. was supported by the NSF CMMT program under Grant No. DMR-2419871 and the Simons Collaboration on Ultra-Quantum Matter, which is a grant from the Simons Foundation (Grant No. 651440).
\end{acknowledgments}

\appendix

\section{Phase freedom in VQE}\label{app:phase}
As discussed in the main text in Section~\ref{sec:widths}, in the variational Wannier state $\ket{\psi_x(\pmb{\theta}^*)}$ obtained with VQE each optimized momentum channel $k$ enters with free phase $\phi_k$ (see Eq.~\eqref{eq:vqe_output}). Here, we examine the distribution of these phases, and their impact on the quasiparticle preparation more closely, focusing on the magnon excitations of the paramagnetic phase. 

\begin{figure}[h!]
    \centering
    \includegraphics[width=0.9\columnwidth]{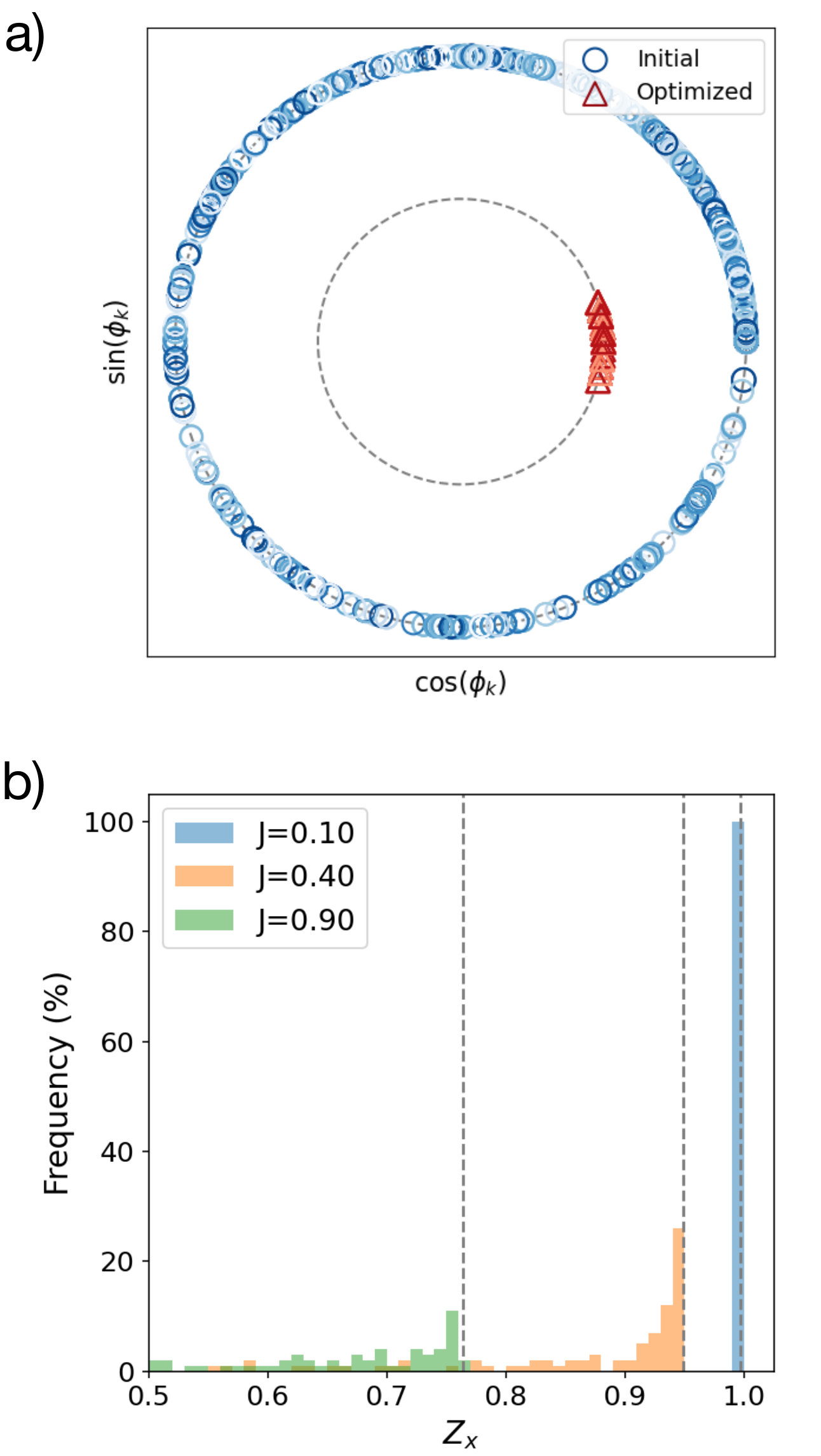}
    \caption{\justifying{\textbf{Phase freedom in the localized quasiparticle states in VQE.} (a) Distribution of phases $\phi_k$ extracted from $\ket{\psi_{x}(\pmb{\theta})}$, before (blue) and after (red) performing the energy minimization to optimize the parameters $\pmb{\theta}$. While circuits with random, not optimized parameters result in uniform phase distribution (blue), the optimization protocol of VQE results in a much narrower distribution, favoring more localized quasiparticle states.  We used $J/h=0.3$. (b) Distribution of quasiparticle weights $Z_x$ for various ratios $J/h$ in the paramagnetic phase. The distribution is skewed towards larger weights, but gets wider upon approaching the critical point $J/h=1$. The dotted lines shows the maximal quasiparticle weights $Z_x^{\rm max}$ evaluated using exact diagonalization.}}
    \label{fig:k-phases}
\end{figure}

We first examine the effect of the optimization protocol in VQE on the distribution of the phases $\phi_k$. Our results are illustrated in Fig. \ref{fig:k-phases}a, where we compare the phase distribution extracted from  $\ket{\psi_x(\pmb{\theta})}$ for a random initial parameter set $\pmb{\theta}_{\rm in}$ (blue), and after the optimization process for the converged parameters  $\pmb{\theta}^*$ (red). Both distributions were obtained by collecting the statistics over multiple  VQE runs,  using different circuit initializations $\pmb{\theta}_{\rm in}$. For a completely random parameter set $\pmb{\theta}_{\rm in}$, we find that the relative phases of the different momentum channels are spread out almost uniformly over the whole range $[0,2\pi]$. However, our results show that the energy minimization procedure alters the distribution drastically, and favors small relative phases between the $k$-channels, yielding a distribution sharply peaked in the vicinity of zero relative phase. We observe the same qualitative behavior across the whole paramagnetic regime, however, the distribution, while still skewed towards small relative phases, becomes more spread out as we approach the phase boundary. As noted in the main text, we attribute these observations to our choice of initial state ~\eqref{eq:initial}, corresponding to the perfectly localized bare spin flip excitation of the limit $J=0$, with all phases $\phi_k$ aligned. This initialization introduces a bias towards more localized Wannier states with small relative phases, which is especially effective deep in the paramagnetic phase, where the quasiparticle states are prepared with shallow circuits. To confirm this intuition, we have verified numerically that modifying the initial state $\ket{\psi^0_x}$ to include random relative phases between momentum channels results in a uniform phase distribution even at the optimized parameter set  $\pmb{\theta}^*$.

In the main text, we discussed that the quasiparticle weight $Z_x$ is a convenient measure of how spread out the variational state $\ket{\psi_x(\pmb{\theta}^*)}$ is, with the maximal weight  $Z_x^{\rm max}$ occurring for perfectly aligned phases $\phi_k$. Comparing the resulting maximally localized Wannier function against the bare spin flip excitation of the limit $J=0$ yields valuable information on the renormalization by interactions. In contrast,  as the different $k$-channels accumulate varying phases, $Z_x$ is reduced and the quasiparticle spreads out, with the shape of the localized quasiparticle state no longer carrying direct information on interaction effects. We examine the histogram of quasiparticle weights  $Z_x$, collected by performing a sufficiently large number of VQE runs with different circuit initializations $\pmb{\theta}_{\rm in}$, in Fig. \ref{fig:k-phases}b. In accordance with the phase distribution peaked around zero relative phase shown in Fig. \ref{fig:k-phases}a, we find that the distribution of $Z_x$ remains skewed towards the maximal weight $Z_x^{\rm max}$ in the whole paramagnetic regime, though the distribution gets broader with increasing ratio $J/h$, as displayed in Fig. \ref{fig:k-phases}b.  We note, however, that the distribution can in principle spread all the way to zero quasiparticle weight, and we indeed observe this behavior, especially for larger ratios $J/h$ closer to the critical point. This skewed distribution of $Z_x$ allows us to find the maximally localized Wannier state via post-selection from a relatively small number of VQE runs for the system sizes considered in this paper, as demonstrated by our results presented in the main text. However, we note that these distributions have to be examined for larger system sizes $N$, to determine the feasibility of such protocols for practical applications targeting moderate to large system sizes. We leave a more detailed investigation for future work.

\section{Spectrum of twisted TFIM}
\label{app:spectrum}

\begin{figure}[t!]
    \centering
    \includegraphics[width=\columnwidth]{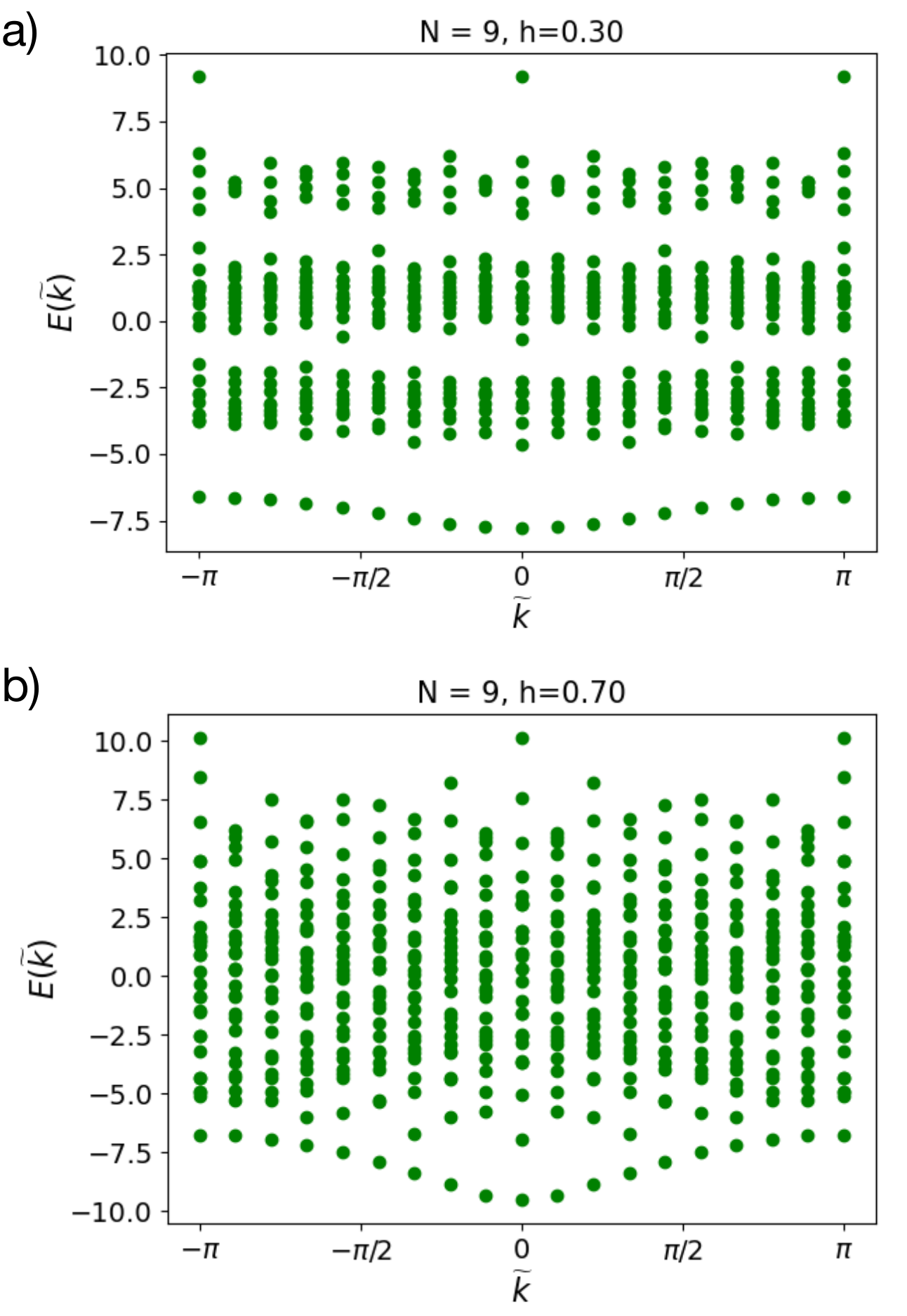}
     \caption{\justifying{\textbf{Full spectrum of the twisted TFIM in the ferromagnetic phase.} Exact diagonalization results for the full spectrum of Hamiltonian ~\eqref{eq:twist}, evaluated for two different ratios $J/h$ within the ferromagnetic regime. The lowest energy states for each wave number $\tilde{k}$ form the single soliton band studied in the main text. We used system size $N=9$.}}
    \label{fig:ED_spectrum}
\end{figure}

In Sec.~\ref{sec:dw}, we introduced a twisted TFIM, Eq.~\eqref{eq:twist}, that allowed us to study single soliton quasiparticles in the ferromagnetic regime, by targeting the lowest energy band of the twisted Hamiltonian. For completeness, here we present the entire energy spectrum of $H_{\rm tTFIM}$, Eq.~\eqref{eq:twist}.

The full spectrum of the twisted model, calculated with exact diagonalization, is shown in Fig.~\ref{fig:ED_spectrum}, for two different ratios $J/h$ within the ferromagnetic regime. As noted above, the lowest energy band captures the single soliton quasiparticles of the original untwisted model, here stabilized by the single antiferromagnetic bond introduced in $H_{\rm tTFIM}$. This soliton band becomes more dispersive with increasing ratio $h/J$, as the solitons become less localized due to the hopping induced by the transverse field $h$. The rest of the energy spectrum is more complex, with higher energy bands reflecting interactions between multiple soliton states. 

\section{Details of VQE optimization in numerical simulations}\label{app:numerics_vqe}

Our numerical simulations of the VQE protocol were performed in Python. The optimization of variational parameters $\pmb{\theta}$ was implemented  using the default BFGS optimizer from \texttt{scipy.optimize.minimize}. We note that the results presented in this work are not significantly affected by  the specific choice of optimization method, as our numerical simulations focus on small system sizes with relatively simple cost landscapes. However, for larger systems in practical quantum simulations, careful selection of the optimization algorithm would be crucial to ensure both efficiency and reliable convergence.

\bibliography{references}

\begin{thebibliography}{40}%
\makeatletter
\providecommand \@ifxundefined [1]{%
 \@ifx{#1\undefined}
}%
\providecommand \@ifnum [1]{%
 \ifnum #1\expandafter \@firstoftwo
 \else \expandafter \@secondoftwo
 \fi
}%
\providecommand \@ifx [1]{%
 \ifx #1\expandafter \@firstoftwo
 \else \expandafter \@secondoftwo
 \fi
}%
\providecommand \natexlab [1]{#1}%
\providecommand \enquote  [1]{``#1''}%
\providecommand \bibnamefont  [1]{#1}%
\providecommand \bibfnamefont [1]{#1}%
\providecommand \citenamefont [1]{#1}%
\providecommand \href@noop [0]{\@secondoftwo}%
\providecommand \href [0]{\begingroup \@sanitize@url \@href}%
\providecommand \@href[1]{\@@startlink{#1}\@@href}%
\providecommand \@@href[1]{\endgroup#1\@@endlink}%
\providecommand \@sanitize@url [0]{\catcode `\\12\catcode `\$12\catcode `\&12\catcode `\#12\catcode `\^12\catcode `\_12\catcode `\%12\relax}%
\providecommand \@@startlink[1]{}%
\providecommand \@@endlink[0]{}%
\providecommand \url  [0]{\begingroup\@sanitize@url \@url }%
\providecommand \@url [1]{\endgroup\@href {#1}{\urlprefix }}%
\providecommand \urlprefix  [0]{URL }%
\providecommand \Eprint [0]{\href }%
\providecommand \doibase [0]{https://doi.org/}%
\providecommand \selectlanguage [0]{\@gobble}%
\providecommand \bibinfo  [0]{\@secondoftwo}%
\providecommand \bibfield  [0]{\@secondoftwo}%
\providecommand \translation [1]{[#1]}%
\providecommand \BibitemOpen [0]{}%
\providecommand \bibitemStop [0]{}%
\providecommand \bibitemNoStop [0]{.\EOS\space}%
\providecommand \EOS [0]{\spacefactor3000\relax}%
\providecommand \BibitemShut  [1]{\csname bibitem#1\endcsname}%
\let\auto@bib@innerbib\@empty
\bibitem [{\citenamefont {Bloch}\ \emph {et~al.}(2008)\citenamefont {Bloch}, \citenamefont {Dalibard},\ and\ \citenamefont {Zwerger}}]{Bloch_RevModPhys.80.885}%
  \BibitemOpen
  \bibfield  {author} {\bibinfo {author} {\bibfnamefont {I.}~\bibnamefont {Bloch}}, \bibinfo {author} {\bibfnamefont {J.}~\bibnamefont {Dalibard}},\ and\ \bibinfo {author} {\bibfnamefont {W.}~\bibnamefont {Zwerger}},\ }\bibfield  {title} {\bibinfo {title} {Many-body physics with ultracold gases},\ }\href {https://doi.org/10.1103/RevModPhys.80.885} {\bibfield  {journal} {\bibinfo  {journal} {Rev. Mod. Phys.}\ }\textbf {\bibinfo {volume} {80}},\ \bibinfo {pages} {885} (\bibinfo {year} {2008})}\BibitemShut {NoStop}%
\bibitem [{\citenamefont {Bernien}\ \emph {et~al.}(2017)\citenamefont {Bernien}, \citenamefont {Schwartz}, \citenamefont {Keesling}, \citenamefont {Levine}, \citenamefont {Omran}, \citenamefont {Pichler}, \citenamefont {Choi}, \citenamefont {Zibrov}, \citenamefont {Endres}, \citenamefont {Greiner}, \citenamefont {Vuleti},\ and\ \citenamefont {Lukin}}]{Bernien2017}%
  \BibitemOpen
  \bibfield  {author} {\bibinfo {author} {\bibfnamefont {H.}~\bibnamefont {Bernien}}, \bibinfo {author} {\bibfnamefont {S.}~\bibnamefont {Schwartz}}, \bibinfo {author} {\bibfnamefont {A.}~\bibnamefont {Keesling}}, \bibinfo {author} {\bibfnamefont {H.}~\bibnamefont {Levine}}, \bibinfo {author} {\bibfnamefont {A.}~\bibnamefont {Omran}}, \bibinfo {author} {\bibfnamefont {H.}~\bibnamefont {Pichler}}, \bibinfo {author} {\bibfnamefont {S.}~\bibnamefont {Choi}}, \bibinfo {author} {\bibfnamefont {A.~S.}\ \bibnamefont {Zibrov}}, \bibinfo {author} {\bibfnamefont {M.}~\bibnamefont {Endres}}, \bibinfo {author} {\bibfnamefont {M.}~\bibnamefont {Greiner}}, \bibinfo {author} {\bibfnamefont {V.}~\bibnamefont {Vuleti}},\ and\ \bibinfo {author} {\bibfnamefont {M.~D.}\ \bibnamefont {Lukin}},\ }\bibfield  {title} {\bibinfo {title} {Probing many-body dynamics on a 51-atom quantum simulator},\ }\href {https://doi.org/10.1038/nature24622} {\bibfield  {journal} {\bibinfo  {journal} {Nature}\ }\textbf {\bibinfo {volume} {551}},\
  \bibinfo {pages} {579} (\bibinfo {year} {2017})}\BibitemShut {NoStop}%
\bibitem [{\citenamefont {Blatt}\ and\ \citenamefont {Roos}(2012)}]{Blatt2012}%
  \BibitemOpen
  \bibfield  {author} {\bibinfo {author} {\bibfnamefont {R.}~\bibnamefont {Blatt}}\ and\ \bibinfo {author} {\bibfnamefont {C.~F.}\ \bibnamefont {Roos}},\ }\bibfield  {title} {\bibinfo {title} {Quantum simulations with trapped ions},\ }\href {https://doi.org/10.1038/nphys2252} {\bibfield  {journal} {\bibinfo  {journal} {Nature Physics}\ }\textbf {\bibinfo {volume} {8}},\ \bibinfo {pages} {277} (\bibinfo {year} {2012})}\BibitemShut {NoStop}%
\bibitem [{\citenamefont {Islam}\ \emph {et~al.}(2013)\citenamefont {Islam}, \citenamefont {Senko}, \citenamefont {Campbell}, \citenamefont {Korenblit}, \citenamefont {Smith}, \citenamefont {Lee}, \citenamefont {Edwards}, \citenamefont {Wang}, \citenamefont {Freericks},\ and\ \citenamefont {Monroe}}]{trappedion_doi:10.1126/science.1232296}%
  \BibitemOpen
  \bibfield  {author} {\bibinfo {author} {\bibfnamefont {R.}~\bibnamefont {Islam}}, \bibinfo {author} {\bibfnamefont {C.}~\bibnamefont {Senko}}, \bibinfo {author} {\bibfnamefont {W.~C.}\ \bibnamefont {Campbell}}, \bibinfo {author} {\bibfnamefont {S.}~\bibnamefont {Korenblit}}, \bibinfo {author} {\bibfnamefont {J.}~\bibnamefont {Smith}}, \bibinfo {author} {\bibfnamefont {A.}~\bibnamefont {Lee}}, \bibinfo {author} {\bibfnamefont {E.~E.}\ \bibnamefont {Edwards}}, \bibinfo {author} {\bibfnamefont {C.-C.~J.}\ \bibnamefont {Wang}}, \bibinfo {author} {\bibfnamefont {J.~K.}\ \bibnamefont {Freericks}},\ and\ \bibinfo {author} {\bibfnamefont {C.}~\bibnamefont {Monroe}},\ }\bibfield  {title} {\bibinfo {title} {Emergence and frustration of magnetism with variable-range interactions in a quantum simulator},\ }\href {https://doi.org/10.1126/science.1232296} {\bibfield  {journal} {\bibinfo  {journal} {Science}\ }\textbf {\bibinfo {volume} {340}},\ \bibinfo {pages} {583} (\bibinfo {year} {2013})}\BibitemShut {NoStop}%
\bibitem [{\citenamefont {Barends}\ \emph {et~al.}(2013)\citenamefont {Barends}, \citenamefont {Kelly}, \citenamefont {Megrant}, \citenamefont {Sank}, \citenamefont {Jeffrey}, \citenamefont {Chen}, \citenamefont {Yin}, \citenamefont {Chiaro}, \citenamefont {Mutus}, \citenamefont {Neill}, \citenamefont {O'Malley}, \citenamefont {Roushan}, \citenamefont {Wenner}, \citenamefont {White}, \citenamefont {Cleland},\ and\ \citenamefont {Martinis}}]{SCqubit_PhysRevLett.111.080502}%
  \BibitemOpen
  \bibfield  {author} {\bibinfo {author} {\bibfnamefont {R.}~\bibnamefont {Barends}}, \bibinfo {author} {\bibfnamefont {J.}~\bibnamefont {Kelly}}, \bibinfo {author} {\bibfnamefont {A.}~\bibnamefont {Megrant}}, \bibinfo {author} {\bibfnamefont {D.}~\bibnamefont {Sank}}, \bibinfo {author} {\bibfnamefont {E.}~\bibnamefont {Jeffrey}}, \bibinfo {author} {\bibfnamefont {Y.}~\bibnamefont {Chen}}, \bibinfo {author} {\bibfnamefont {Y.}~\bibnamefont {Yin}}, \bibinfo {author} {\bibfnamefont {B.}~\bibnamefont {Chiaro}}, \bibinfo {author} {\bibfnamefont {J.}~\bibnamefont {Mutus}}, \bibinfo {author} {\bibfnamefont {C.}~\bibnamefont {Neill}}, \bibinfo {author} {\bibfnamefont {P.}~\bibnamefont {O'Malley}}, \bibinfo {author} {\bibfnamefont {P.}~\bibnamefont {Roushan}}, \bibinfo {author} {\bibfnamefont {J.}~\bibnamefont {Wenner}}, \bibinfo {author} {\bibfnamefont {T.~C.}\ \bibnamefont {White}}, \bibinfo {author} {\bibfnamefont {A.~N.}\ \bibnamefont {Cleland}},\ and\ \bibinfo {author} {\bibfnamefont {J.~M.}\ \bibnamefont
  {Martinis}},\ }\bibfield  {title} {\bibinfo {title} {Coherent josephson qubit suitable for scalable quantum integrated circuits},\ }\href {https://doi.org/10.1103/PhysRevLett.111.080502} {\bibfield  {journal} {\bibinfo  {journal} {Phys. Rev. Lett.}\ }\textbf {\bibinfo {volume} {111}},\ \bibinfo {pages} {080502} (\bibinfo {year} {2013})}\BibitemShut {NoStop}%
\bibitem [{\citenamefont {Gambetta}\ \emph {et~al.}(2017)\citenamefont {Gambetta}, \citenamefont {Chow},\ and\ \citenamefont {Steffen}}]{Gambetta2017}%
  \BibitemOpen
  \bibfield  {author} {\bibinfo {author} {\bibfnamefont {J.~M.}\ \bibnamefont {Gambetta}}, \bibinfo {author} {\bibfnamefont {J.~M.}\ \bibnamefont {Chow}},\ and\ \bibinfo {author} {\bibfnamefont {M.}~\bibnamefont {Steffen}},\ }\bibfield  {title} {\bibinfo {title} {Building logical qubits in a superconducting quantum computing system},\ }\href {https://doi.org/10.1038/s41534-016-0004-0} {\bibfield  {journal} {\bibinfo  {journal} {npj Quantum Information}\ }\textbf {\bibinfo {volume} {3}},\ \bibinfo {pages} {2} (\bibinfo {year} {2017})}\BibitemShut {NoStop}%
\bibitem [{\citenamefont {Giovannetti}\ \emph {et~al.}(2011)\citenamefont {Giovannetti}, \citenamefont {Lloyd},\ and\ \citenamefont {Maccone}}]{Giovannetti2011}%
  \BibitemOpen
  \bibfield  {author} {\bibinfo {author} {\bibfnamefont {V.}~\bibnamefont {Giovannetti}}, \bibinfo {author} {\bibfnamefont {S.}~\bibnamefont {Lloyd}},\ and\ \bibinfo {author} {\bibfnamefont {L.}~\bibnamefont {Maccone}},\ }\bibfield  {title} {\bibinfo {title} {Advances in quantum metrology},\ }\href {https://doi.org/10.1038/nphoton.2011.35} {\bibfield  {journal} {\bibinfo  {journal} {Nature Photonics}\ }\textbf {\bibinfo {volume} {5}},\ \bibinfo {pages} {222} (\bibinfo {year} {2011})}\BibitemShut {NoStop}%
\bibitem [{\citenamefont {Degen}\ \emph {et~al.}(2017)\citenamefont {Degen}, \citenamefont {Reinhard},\ and\ \citenamefont {Cappellaro}}]{RevModPhys.89.035002}%
  \BibitemOpen
  \bibfield  {author} {\bibinfo {author} {\bibfnamefont {C.~L.}\ \bibnamefont {Degen}}, \bibinfo {author} {\bibfnamefont {F.}~\bibnamefont {Reinhard}},\ and\ \bibinfo {author} {\bibfnamefont {P.}~\bibnamefont {Cappellaro}},\ }\bibfield  {title} {\bibinfo {title} {Quantum sensing},\ }\href {https://doi.org/10.1103/RevModPhys.89.035002} {\bibfield  {journal} {\bibinfo  {journal} {Rev. Mod. Phys.}\ }\textbf {\bibinfo {volume} {89}},\ \bibinfo {pages} {035002} (\bibinfo {year} {2017})}\BibitemShut {NoStop}%
\bibitem [{\citenamefont {Preskill}(2018)}]{Preskill2018quantumcomputingin}%
  \BibitemOpen
  \bibfield  {author} {\bibinfo {author} {\bibfnamefont {J.}~\bibnamefont {Preskill}},\ }\bibfield  {title} {\bibinfo {title} {Quantum {C}omputing in the {NISQ} era and beyond},\ }\href {https://doi.org/10.22331/q-2018-08-06-79} {\bibfield  {journal} {\bibinfo  {journal} {{Quantum}}\ }\textbf {\bibinfo {volume} {2}},\ \bibinfo {pages} {79} (\bibinfo {year} {2018})}\BibitemShut {NoStop}%
\bibitem [{\citenamefont {Peruzzo}\ \emph {et~al.}(2014)\citenamefont {Peruzzo}, \citenamefont {McClean}, \citenamefont {Shadbolt}, \citenamefont {Yung}, \citenamefont {Zhou}, \citenamefont {Love}, \citenamefont {Aspuru-Guzik},\ and\ \citenamefont {O{\^a}Brien}}]{Peruzzo2014}%
  \BibitemOpen
  \bibfield  {author} {\bibinfo {author} {\bibfnamefont {A.}~\bibnamefont {Peruzzo}}, \bibinfo {author} {\bibfnamefont {J.}~\bibnamefont {McClean}}, \bibinfo {author} {\bibfnamefont {P.}~\bibnamefont {Shadbolt}}, \bibinfo {author} {\bibfnamefont {M.-H.}\ \bibnamefont {Yung}}, \bibinfo {author} {\bibfnamefont {X.-Q.}\ \bibnamefont {Zhou}}, \bibinfo {author} {\bibfnamefont {P.~J.}\ \bibnamefont {Love}}, \bibinfo {author} {\bibfnamefont {A.}~\bibnamefont {Aspuru-Guzik}},\ and\ \bibinfo {author} {\bibfnamefont {J.~L.}\ \bibnamefont {O{\^a}Brien}},\ }\bibfield  {title} {\bibinfo {title} {A variational eigenvalue solver on a photonic quantum processor},\ }\href {https://doi.org/10.1038/ncomms5213} {\bibfield  {journal} {\bibinfo  {journal} {Nature Communications}\ }\textbf {\bibinfo {volume} {5}},\ \bibinfo {pages} {4213} (\bibinfo {year} {2014})}\BibitemShut {NoStop}%
\bibitem [{\citenamefont {Bauer}\ \emph {et~al.}(2016)\citenamefont {Bauer}, \citenamefont {Wecker}, \citenamefont {Millis}, \citenamefont {Hastings},\ and\ \citenamefont {Troyer}}]{VQE_PhysRevX.6.031045}%
  \BibitemOpen
  \bibfield  {author} {\bibinfo {author} {\bibfnamefont {B.}~\bibnamefont {Bauer}}, \bibinfo {author} {\bibfnamefont {D.}~\bibnamefont {Wecker}}, \bibinfo {author} {\bibfnamefont {A.~J.}\ \bibnamefont {Millis}}, \bibinfo {author} {\bibfnamefont {M.~B.}\ \bibnamefont {Hastings}},\ and\ \bibinfo {author} {\bibfnamefont {M.}~\bibnamefont {Troyer}},\ }\bibfield  {title} {\bibinfo {title} {Hybrid quantum-classical approach to correlated materials},\ }\href {https://doi.org/10.1103/PhysRevX.6.031045} {\bibfield  {journal} {\bibinfo  {journal} {Phys. Rev. X}\ }\textbf {\bibinfo {volume} {6}},\ \bibinfo {pages} {031045} (\bibinfo {year} {2016})}\BibitemShut {NoStop}%
\bibitem [{\citenamefont {Kandala}\ \emph {et~al.}(2017)\citenamefont {Kandala}, \citenamefont {Mezzacapo}, \citenamefont {Temme}, \citenamefont {Takita}, \citenamefont {Brink}, \citenamefont {Chow},\ and\ \citenamefont {Gambetta}}]{Kandala2017}%
  \BibitemOpen
  \bibfield  {author} {\bibinfo {author} {\bibfnamefont {A.}~\bibnamefont {Kandala}}, \bibinfo {author} {\bibfnamefont {A.}~\bibnamefont {Mezzacapo}}, \bibinfo {author} {\bibfnamefont {K.}~\bibnamefont {Temme}}, \bibinfo {author} {\bibfnamefont {M.}~\bibnamefont {Takita}}, \bibinfo {author} {\bibfnamefont {M.}~\bibnamefont {Brink}}, \bibinfo {author} {\bibfnamefont {J.~M.}\ \bibnamefont {Chow}},\ and\ \bibinfo {author} {\bibfnamefont {J.~M.}\ \bibnamefont {Gambetta}},\ }\bibfield  {title} {\bibinfo {title} {Hardware-efficient variational quantum eigensolver for small molecules and quantum magnets},\ }\href {https://doi.org/10.1038/nature23879} {\bibfield  {journal} {\bibinfo  {journal} {Nature}\ }\textbf {\bibinfo {volume} {549}},\ \bibinfo {pages} {242} (\bibinfo {year} {2017})}\BibitemShut {NoStop}%
\bibitem [{\citenamefont {Ho}\ and\ \citenamefont {Hsieh}(2019)}]{SciPostPhys.6.3.029}%
  \BibitemOpen
  \bibfield  {author} {\bibinfo {author} {\bibfnamefont {W.~W.}\ \bibnamefont {Ho}}\ and\ \bibinfo {author} {\bibfnamefont {T.~H.}\ \bibnamefont {Hsieh}},\ }\bibfield  {title} {\bibinfo {title} {{Efficient variational simulation of non-trivial quantum states}},\ }\href {https://doi.org/10.21468/SciPostPhys.6.3.029} {\bibfield  {journal} {\bibinfo  {journal} {SciPost Phys.}\ }\textbf {\bibinfo {volume} {6}},\ \bibinfo {pages} {029} (\bibinfo {year} {2019})}\BibitemShut {NoStop}%
\bibitem [{\citenamefont {Tilly}\ \emph {et~al.}(2022)\citenamefont {Tilly}, \citenamefont {Chen}, \citenamefont {Cao}, \citenamefont {Picozzi}, \citenamefont {Setia}, \citenamefont {Li}, \citenamefont {Grant}, \citenamefont {Wossnig}, \citenamefont {Rungger}, \citenamefont {Booth},\ and\ \citenamefont {Tennyson}}]{TILLY20221}%
  \BibitemOpen
  \bibfield  {author} {\bibinfo {author} {\bibfnamefont {J.}~\bibnamefont {Tilly}}, \bibinfo {author} {\bibfnamefont {H.}~\bibnamefont {Chen}}, \bibinfo {author} {\bibfnamefont {S.}~\bibnamefont {Cao}}, \bibinfo {author} {\bibfnamefont {D.}~\bibnamefont {Picozzi}}, \bibinfo {author} {\bibfnamefont {K.}~\bibnamefont {Setia}}, \bibinfo {author} {\bibfnamefont {Y.}~\bibnamefont {Li}}, \bibinfo {author} {\bibfnamefont {E.}~\bibnamefont {Grant}}, \bibinfo {author} {\bibfnamefont {L.}~\bibnamefont {Wossnig}}, \bibinfo {author} {\bibfnamefont {I.}~\bibnamefont {Rungger}}, \bibinfo {author} {\bibfnamefont {G.~H.}\ \bibnamefont {Booth}},\ and\ \bibinfo {author} {\bibfnamefont {J.}~\bibnamefont {Tennyson}},\ }\bibfield  {title} {\bibinfo {title} {The variational quantum eigensolver: A review of methods and best practices},\ }\href {https://doi.org/https://doi.org/10.1016/j.physrep.2022.08.003} {\bibfield  {journal} {\bibinfo  {journal} {Physics Reports}\ }\textbf {\bibinfo {volume} {986}},\ \bibinfo {pages} {1}
  (\bibinfo {year} {2022})},\ \bibinfo {note} {the Variational Quantum Eigensolver: a review of methods and best practices}\BibitemShut {NoStop}%
\bibitem [{\citenamefont {Nakanishi}\ \emph {et~al.}(2019)\citenamefont {Nakanishi}, \citenamefont {Mitarai},\ and\ \citenamefont {Fujii}}]{excited_PhysRevResearch.1.033062}%
  \BibitemOpen
  \bibfield  {author} {\bibinfo {author} {\bibfnamefont {K.~M.}\ \bibnamefont {Nakanishi}}, \bibinfo {author} {\bibfnamefont {K.}~\bibnamefont {Mitarai}},\ and\ \bibinfo {author} {\bibfnamefont {K.}~\bibnamefont {Fujii}},\ }\bibfield  {title} {\bibinfo {title} {Subspace-search variational quantum eigensolver for excited states},\ }\href {https://doi.org/10.1103/PhysRevResearch.1.033062} {\bibfield  {journal} {\bibinfo  {journal} {Phys. Rev. Res.}\ }\textbf {\bibinfo {volume} {1}},\ \bibinfo {pages} {033062} (\bibinfo {year} {2019})}\BibitemShut {NoStop}%
\bibitem [{\citenamefont {Wu}\ and\ \citenamefont {Hsieh}(2019)}]{thermal_PhysRevLett.123.220502}%
  \BibitemOpen
  \bibfield  {author} {\bibinfo {author} {\bibfnamefont {J.}~\bibnamefont {Wu}}\ and\ \bibinfo {author} {\bibfnamefont {T.~H.}\ \bibnamefont {Hsieh}},\ }\bibfield  {title} {\bibinfo {title} {Variational thermal quantum simulation via thermofield double states},\ }\href {https://doi.org/10.1103/PhysRevLett.123.220502} {\bibfield  {journal} {\bibinfo  {journal} {Phys. Rev. Lett.}\ }\textbf {\bibinfo {volume} {123}},\ \bibinfo {pages} {220502} (\bibinfo {year} {2019})}\BibitemShut {NoStop}%
\bibitem [{\citenamefont {McClean}\ \emph {et~al.}(2016)\citenamefont {McClean}, \citenamefont {Romero}, \citenamefont {Babbush},\ and\ \citenamefont {Aspuru-Guzik}}]{McClean_2016}%
  \BibitemOpen
  \bibfield  {author} {\bibinfo {author} {\bibfnamefont {J.~R.}\ \bibnamefont {McClean}}, \bibinfo {author} {\bibfnamefont {J.}~\bibnamefont {Romero}}, \bibinfo {author} {\bibfnamefont {R.}~\bibnamefont {Babbush}},\ and\ \bibinfo {author} {\bibfnamefont {A.}~\bibnamefont {Aspuru-Guzik}},\ }\bibfield  {title} {\bibinfo {title} {The theory of variational hybrid quantum-classical algorithms},\ }\href {https://doi.org/10.1088/1367-2630/18/2/023023} {\bibfield  {journal} {\bibinfo  {journal} {New Journal of Physics}\ }\textbf {\bibinfo {volume} {18}},\ \bibinfo {pages} {023023} (\bibinfo {year} {2016})}\BibitemShut {NoStop}%
\bibitem [{\citenamefont {Higgott}\ \emph {et~al.}(2019)\citenamefont {Higgott}, \citenamefont {Wang},\ and\ \citenamefont {Brierley}}]{Higgott2019variationalquantum}%
  \BibitemOpen
  \bibfield  {author} {\bibinfo {author} {\bibfnamefont {O.}~\bibnamefont {Higgott}}, \bibinfo {author} {\bibfnamefont {D.}~\bibnamefont {Wang}},\ and\ \bibinfo {author} {\bibfnamefont {S.}~\bibnamefont {Brierley}},\ }\bibfield  {title} {\bibinfo {title} {Variational {Q}uantum {C}omputation of {E}xcited {S}tates},\ }\href {https://doi.org/10.22331/q-2019-07-01-156} {\bibfield  {journal} {\bibinfo  {journal} {{Quantum}}\ }\textbf {\bibinfo {volume} {3}},\ \bibinfo {pages} {156} (\bibinfo {year} {2019})}\BibitemShut {NoStop}%
\bibitem [{\citenamefont {Abrikosov}\ \emph {et~al.}(1975)\citenamefont {Abrikosov}, \citenamefont {Dzyaloshinskii},\ and\ \citenamefont {Gorkov}}]{Abrikosov:107441}%
  \BibitemOpen
  \bibfield  {author} {\bibinfo {author} {\bibfnamefont {A.~A.}\ \bibnamefont {Abrikosov}}, \bibinfo {author} {\bibfnamefont {I.}~\bibnamefont {Dzyaloshinskii}},\ and\ \bibinfo {author} {\bibfnamefont {L.~P.}\ \bibnamefont {Gorkov}},\ }\href@noop {} {\emph {\bibinfo {title} {{Methods of quantum field theory in statistical physics}}}}\ (\bibinfo  {publisher} {Dover Publications},\ \bibinfo {address} {New York, NY},\ \bibinfo {year} {1975})\BibitemShut {NoStop}%
\bibitem [{\citenamefont {Negele}\ and\ \citenamefont {Orland}(1988)}]{NegeleOrland}%
  \BibitemOpen
  \bibfield  {author} {\bibinfo {author} {\bibfnamefont {J.~W.}\ \bibnamefont {Negele}}\ and\ \bibinfo {author} {\bibfnamefont {H.}~\bibnamefont {Orland}},\ }\href@noop {} {\emph {\bibinfo {title} {Quantum many-particle systems}}}\ (\bibinfo  {publisher} {Westview Press},\ \bibinfo {address} {Boulder},\ \bibinfo {year} {1988})\BibitemShut {NoStop}%
\bibitem [{\citenamefont {Coleman}(2015)}]{Coleman_2015}%
  \BibitemOpen
  \bibfield  {author} {\bibinfo {author} {\bibfnamefont {P.}~\bibnamefont {Coleman}},\ }\href@noop {} {\emph {\bibinfo {title} {Introduction to Many-Body Physics}}}\ (\bibinfo  {publisher} {Cambridge University Press},\ \bibinfo {year} {2015})\BibitemShut {NoStop}%
\bibitem [{\citenamefont {Pfeuty}(1970)}]{pfeuty1970}%
  \BibitemOpen
  \bibfield  {author} {\bibinfo {author} {\bibfnamefont {P.}~\bibnamefont {Pfeuty}},\ }\bibfield  {title} {\bibinfo {title} {The one-dimensional ising model with a transverse field},\ }\href@noop {} {\bibfield  {journal} {\bibinfo  {journal} {ANNALS of Physics}\ }\textbf {\bibinfo {volume} {57}},\ \bibinfo {pages} {79} (\bibinfo {year} {1970})}\BibitemShut {NoStop}%
\bibitem [{\citenamefont {Mikeska}\ and\ \citenamefont {Miyashita}(1996)}]{Mikeska1996}%
  \BibitemOpen
  \bibfield  {author} {\bibinfo {author} {\bibfnamefont {H.-J.}\ \bibnamefont {Mikeska}}\ and\ \bibinfo {author} {\bibfnamefont {S.}~\bibnamefont {Miyashita}},\ }\bibfield  {title} {\bibinfo {title} {Classial and quantum solitons in the transverse field ising model},\ }\href {https://doi.org/10.1007/s002570050209} {\bibfield  {journal} {\bibinfo  {journal} {Zeitschrift f{\"u}r Physik B Condensed Matter}\ }\textbf {\bibinfo {volume} {101}},\ \bibinfo {pages} {275} (\bibinfo {year} {1996})}\BibitemShut {NoStop}%
\bibitem [{\citenamefont {Jones}\ \emph {et~al.}(2019)\citenamefont {Jones}, \citenamefont {Endo}, \citenamefont {McArdle}, \citenamefont {Yuan},\ and\ \citenamefont {Benjamin}}]{PhysRevA.99.062304}%
  \BibitemOpen
  \bibfield  {author} {\bibinfo {author} {\bibfnamefont {T.}~\bibnamefont {Jones}}, \bibinfo {author} {\bibfnamefont {S.}~\bibnamefont {Endo}}, \bibinfo {author} {\bibfnamefont {S.}~\bibnamefont {McArdle}}, \bibinfo {author} {\bibfnamefont {X.}~\bibnamefont {Yuan}},\ and\ \bibinfo {author} {\bibfnamefont {S.~C.}\ \bibnamefont {Benjamin}},\ }\bibfield  {title} {\bibinfo {title} {Variational quantum algorithms for discovering hamiltonian spectra},\ }\href {https://doi.org/10.1103/PhysRevA.99.062304} {\bibfield  {journal} {\bibinfo  {journal} {Phys. Rev. A}\ }\textbf {\bibinfo {volume} {99}},\ \bibinfo {pages} {062304} (\bibinfo {year} {2019})}\BibitemShut {NoStop}%
\bibitem [{\citenamefont {Gard}\ \emph {et~al.}(2020)\citenamefont {Gard}, \citenamefont {Zhu}, \citenamefont {Barron}, \citenamefont {Mayhall}, \citenamefont {Economou},\ and\ \citenamefont {Barnes}}]{Gard2020}%
  \BibitemOpen
  \bibfield  {author} {\bibinfo {author} {\bibfnamefont {B.~T.}\ \bibnamefont {Gard}}, \bibinfo {author} {\bibfnamefont {L.}~\bibnamefont {Zhu}}, \bibinfo {author} {\bibfnamefont {G.~S.}\ \bibnamefont {Barron}}, \bibinfo {author} {\bibfnamefont {N.~J.}\ \bibnamefont {Mayhall}}, \bibinfo {author} {\bibfnamefont {S.~E.}\ \bibnamefont {Economou}},\ and\ \bibinfo {author} {\bibfnamefont {E.}~\bibnamefont {Barnes}},\ }\bibfield  {title} {\bibinfo {title} {Efficient symmetry-preserving state preparation circuits for the variational quantum eigensolver algorithm},\ }\href {https://doi.org/10.1038/s41534-019-0240-1} {\bibfield  {journal} {\bibinfo  {journal} {npj Quantum Information}\ }\textbf {\bibinfo {volume} {6}},\ \bibinfo {pages} {10} (\bibinfo {year} {2020})}\BibitemShut {NoStop}%
\bibitem [{\citenamefont {Ryabinkin}\ \emph {et~al.}(2019)\citenamefont {Ryabinkin}, \citenamefont {Genin},\ and\ \citenamefont {Izmaylov}}]{Ryabinkin2019}%
  \BibitemOpen
  \bibfield  {author} {\bibinfo {author} {\bibfnamefont {I.~G.}\ \bibnamefont {Ryabinkin}}, \bibinfo {author} {\bibfnamefont {S.~N.}\ \bibnamefont {Genin}},\ and\ \bibinfo {author} {\bibfnamefont {A.~F.}\ \bibnamefont {Izmaylov}},\ }\bibfield  {title} {\bibinfo {title} {Constrained variational quantum eigensolver: Quantum computer search engine in the fock space},\ }\href {https://doi.org/10.1021/acs.jctc.8b00943} {\bibfield  {journal} {\bibinfo  {journal} {Journal of Chemical Theory and Computation}\ }\textbf {\bibinfo {volume} {15}},\ \bibinfo {pages} {249} (\bibinfo {year} {2019})}\BibitemShut {NoStop}%
\bibitem [{\citenamefont {Seki}\ \emph {et~al.}(2020)\citenamefont {Seki}, \citenamefont {Shirakawa},\ and\ \citenamefont {Yunoki}}]{symm_adapt_PhysRevA.101.052340}%
  \BibitemOpen
  \bibfield  {author} {\bibinfo {author} {\bibfnamefont {K.}~\bibnamefont {Seki}}, \bibinfo {author} {\bibfnamefont {T.}~\bibnamefont {Shirakawa}},\ and\ \bibinfo {author} {\bibfnamefont {S.}~\bibnamefont {Yunoki}},\ }\bibfield  {title} {\bibinfo {title} {Symmetry-adapted variational quantum eigensolver},\ }\href {https://doi.org/10.1103/PhysRevA.101.052340} {\bibfield  {journal} {\bibinfo  {journal} {Phys. Rev. A}\ }\textbf {\bibinfo {volume} {101}},\ \bibinfo {pages} {052340} (\bibinfo {year} {2020})}\BibitemShut {NoStop}%
\bibitem [{\citenamefont {Sachdev}(1999)}]{Sachdev1999}%
  \BibitemOpen
  \bibfield  {author} {\bibinfo {author} {\bibfnamefont {S.}~\bibnamefont {Sachdev}},\ }\href@noop {} {\emph {\bibinfo {title} {Quantum Phase Transitions}}}\ (\bibinfo  {publisher} {Cambridge University Press},\ \bibinfo {address} {London},\ \bibinfo {year} {1999})\BibitemShut {NoStop}%
\bibitem [{\citenamefont {Coldea}\ \emph {et~al.}(2010)\citenamefont {Coldea}, \citenamefont {Tennant}, \citenamefont {Wheeler}, \citenamefont {Wawrzynska}, \citenamefont {Prabhakaran}, \citenamefont {Telling}, \citenamefont {Habicht}, \citenamefont {Smeibidl},\ and\ \citenamefont {Kiefer}}]{Coldea2010}%
  \BibitemOpen
  \bibfield  {author} {\bibinfo {author} {\bibfnamefont {R.}~\bibnamefont {Coldea}}, \bibinfo {author} {\bibfnamefont {D.~A.}\ \bibnamefont {Tennant}}, \bibinfo {author} {\bibfnamefont {E.~M.}\ \bibnamefont {Wheeler}}, \bibinfo {author} {\bibfnamefont {E.}~\bibnamefont {Wawrzynska}}, \bibinfo {author} {\bibfnamefont {D.}~\bibnamefont {Prabhakaran}}, \bibinfo {author} {\bibfnamefont {M.}~\bibnamefont {Telling}}, \bibinfo {author} {\bibfnamefont {K.}~\bibnamefont {Habicht}}, \bibinfo {author} {\bibfnamefont {P.}~\bibnamefont {Smeibidl}},\ and\ \bibinfo {author} {\bibfnamefont {K.}~\bibnamefont {Kiefer}},\ }\bibfield  {title} {\bibinfo {title} {Quantum criticality in an {I}sing chain: Experimental evidence for emergent {E8} symmetry},\ }\href {https://doi.org/10.1126/science.1180085} {\bibfield  {journal} {\bibinfo  {journal} {Science}\ }\textbf {\bibinfo {volume} {327}},\ \bibinfo {pages} {177} (\bibinfo {year} {2010})}\BibitemShut {NoStop}%
\bibitem [{\citenamefont {Leibfried}\ \emph {et~al.}(2004)\citenamefont {Leibfried}, \citenamefont {Barrett}, \citenamefont {Schaetz}, \citenamefont {Britton}, \citenamefont {Chiaverini}, \citenamefont {Itano}, \citenamefont {Jost}, \citenamefont {Langer},\ and\ \citenamefont {Wineland}}]{Leibfried2004}%
  \BibitemOpen
  \bibfield  {author} {\bibinfo {author} {\bibfnamefont {D.}~\bibnamefont {Leibfried}}, \bibinfo {author} {\bibfnamefont {M.~D.}\ \bibnamefont {Barrett}}, \bibinfo {author} {\bibfnamefont {T.}~\bibnamefont {Schaetz}}, \bibinfo {author} {\bibfnamefont {J.}~\bibnamefont {Britton}}, \bibinfo {author} {\bibfnamefont {J.}~\bibnamefont {Chiaverini}}, \bibinfo {author} {\bibfnamefont {W.~M.}\ \bibnamefont {Itano}}, \bibinfo {author} {\bibfnamefont {J.~D.}\ \bibnamefont {Jost}}, \bibinfo {author} {\bibfnamefont {C.}~\bibnamefont {Langer}},\ and\ \bibinfo {author} {\bibfnamefont {D.~J.}\ \bibnamefont {Wineland}},\ }\bibfield  {title} {\bibinfo {title} {Toward heisenberg-limited spectroscopy with multiparticle entangled states},\ }\href {https://doi.org/10.1126/science.1097576} {\bibfield  {journal} {\bibinfo  {journal} {Science}\ }\textbf {\bibinfo {volume} {304}},\ \bibinfo {pages} {1476} (\bibinfo {year} {2004})}\BibitemShut {NoStop}%
\bibitem [{Note1()}]{Note1}%
  \BibitemOpen
  \bibinfo {note} {Full translation symmetry is broken by random circuits consisting of local unitary gates, but they can still preserve symmetry against translation by two lattice sites.}\BibitemShut {Stop}%
\bibitem [{\citenamefont {Li}\ and\ \citenamefont {Benjamin}(2017)}]{CU_PhysRevX.7.021050}%
  \BibitemOpen
  \bibfield  {author} {\bibinfo {author} {\bibfnamefont {Y.}~\bibnamefont {Li}}\ and\ \bibinfo {author} {\bibfnamefont {S.~C.}\ \bibnamefont {Benjamin}},\ }\bibfield  {title} {\bibinfo {title} {Efficient variational quantum simulator incorporating active error minimization},\ }\href {https://doi.org/10.1103/PhysRevX.7.021050} {\bibfield  {journal} {\bibinfo  {journal} {Phys. Rev. X}\ }\textbf {\bibinfo {volume} {7}},\ \bibinfo {pages} {021050} (\bibinfo {year} {2017})}\BibitemShut {NoStop}%
\bibitem [{Note2()}]{Note2}%
  \BibitemOpen
  \bibinfo {note} {We note that the symmetry adapted variational protocol proposed in Ref. \cite {symm_adapt_PhysRevA.101.052340} involves a similar procedure, allowing in principle to imprint the desired momentum on the variational wave function prepared in the quantum device. A notable difference compared to our approach is the need to perform this operation in each step of the optimization process, In contrast, by working with Wannier states, we ensure that it has to be performed only once on the already optimized wave function to access the full dispersion relation.}\BibitemShut {Stop}%
\bibitem [{\citenamefont {Childress}\ and\ \citenamefont {Hanson}(2013)}]{Childress2013}%
  \BibitemOpen
  \bibfield  {author} {\bibinfo {author} {\bibfnamefont {L.}~\bibnamefont {Childress}}\ and\ \bibinfo {author} {\bibfnamefont {R.}~\bibnamefont {Hanson}},\ }\bibfield  {title} {\bibinfo {title} {Diamond nv centers for quantum computing and quantum networks},\ }\href {https://doi.org/10.1557/mrs.2013.20} {\bibfield  {journal} {\bibinfo  {journal} {MRS Bulletin}\ }\textbf {\bibinfo {volume} {38}},\ \bibinfo {pages} {134} (\bibinfo {year} {2013})}\BibitemShut {NoStop}%
\bibitem [{\citenamefont {Lee}\ \emph {et~al.}(2017)\citenamefont {Lee}, \citenamefont {Lee}, \citenamefont {Cady}, \citenamefont {Ovartchaiyapong},\ and\ \citenamefont {Jayich}}]{Lee_2017}%
  \BibitemOpen
  \bibfield  {author} {\bibinfo {author} {\bibfnamefont {D.}~\bibnamefont {Lee}}, \bibinfo {author} {\bibfnamefont {K.~W.}\ \bibnamefont {Lee}}, \bibinfo {author} {\bibfnamefont {J.~V.}\ \bibnamefont {Cady}}, \bibinfo {author} {\bibfnamefont {P.}~\bibnamefont {Ovartchaiyapong}},\ and\ \bibinfo {author} {\bibfnamefont {A.~C.~B.}\ \bibnamefont {Jayich}},\ }\bibfield  {title} {\bibinfo {title} {Topical review: spins and mechanics in diamond},\ }\href {https://doi.org/10.1088/2040-8986/aa52cd} {\bibfield  {journal} {\bibinfo  {journal} {Journal of Optics}\ }\textbf {\bibinfo {volume} {19}},\ \bibinfo {pages} {033001} (\bibinfo {year} {2017})}\BibitemShut {NoStop}%
\bibitem [{\citenamefont {Arute}\ \emph {et~al.}(2019)\citenamefont {Arute}, \citenamefont {Arya}, \citenamefont {Babbush}, \citenamefont {Bacon}, \citenamefont {Bardin}, \citenamefont {Barends}, \citenamefont {Biswas}, \citenamefont {Boixo}, \citenamefont {Brandao}, \citenamefont {Buell}, \citenamefont {Burkett}, \citenamefont {Chen}, \citenamefont {Chen}, \citenamefont {Chiaro}, \citenamefont {Collins}, \citenamefont {Courtney}, \citenamefont {Dunsworth}, \citenamefont {Farhi}, \citenamefont {Foxen}, \citenamefont {Fowler}, \citenamefont {Gidney}, \citenamefont {Giustina}, \citenamefont {Graff}, \citenamefont {Guerin}, \citenamefont {Habegger}, \citenamefont {Harrigan}, \citenamefont {Hartmann}, \citenamefont {Ho}, \citenamefont {Hoffmann}, \citenamefont {Huang}, \citenamefont {Humble}, \citenamefont {Isakov}, \citenamefont {Jeffrey}, \citenamefont {Jiang}, \citenamefont {Kafri}, \citenamefont {Kechedzhi}, \citenamefont {Kelly}, \citenamefont {Klimov}, \citenamefont {Knysh}, \citenamefont {Korotkov},
  \citenamefont {Kostritsa}, \citenamefont {Landhuis}, \citenamefont {Lindmark}, \citenamefont {Lucero}, \citenamefont {Lyakh}, \citenamefont {Mandr{\~A} }, \citenamefont {McClean}, \citenamefont {McEwen}, \citenamefont {Megrant}, \citenamefont {Mi}, \citenamefont {Michielsen}, \citenamefont {Mohseni}, \citenamefont {Mutus}, \citenamefont {Naaman}, \citenamefont {Neeley}, \citenamefont {Neill}, \citenamefont {Niu}, \citenamefont {Ostby}, \citenamefont {Petukhov}, \citenamefont {Platt}, \citenamefont {Quintana}, \citenamefont {Rieffel}, \citenamefont {Roushan}, \citenamefont {Rubin}, \citenamefont {Sank}, \citenamefont {Satzinger}, \citenamefont {Smelyanskiy}, \citenamefont {Sung}, \citenamefont {Trevithick}, \citenamefont {Vainsencher}, \citenamefont {Villalonga}, \citenamefont {White}, \citenamefont {Yao}, \citenamefont {Yeh}, \citenamefont {Zalcman}, \citenamefont {Neven},\ and\ \citenamefont {Martinis}}]{Arute2019}%
  \BibitemOpen
  \bibfield  {author} {\bibinfo {author} {\bibfnamefont {F.}~\bibnamefont {Arute}}, \bibinfo {author} {\bibfnamefont {K.}~\bibnamefont {Arya}}, \bibinfo {author} {\bibfnamefont {R.}~\bibnamefont {Babbush}}, \bibinfo {author} {\bibfnamefont {D.}~\bibnamefont {Bacon}}, \bibinfo {author} {\bibfnamefont {J.~C.}\ \bibnamefont {Bardin}}, \bibinfo {author} {\bibfnamefont {R.}~\bibnamefont {Barends}}, \bibinfo {author} {\bibfnamefont {R.}~\bibnamefont {Biswas}}, \bibinfo {author} {\bibfnamefont {S.}~\bibnamefont {Boixo}}, \bibinfo {author} {\bibfnamefont {F.~G. S.~L.}\ \bibnamefont {Brandao}}, \bibinfo {author} {\bibfnamefont {D.~A.}\ \bibnamefont {Buell}}, \bibinfo {author} {\bibfnamefont {B.}~\bibnamefont {Burkett}}, \bibinfo {author} {\bibfnamefont {Y.}~\bibnamefont {Chen}}, \bibinfo {author} {\bibfnamefont {Z.}~\bibnamefont {Chen}}, \bibinfo {author} {\bibfnamefont {B.}~\bibnamefont {Chiaro}}, \bibinfo {author} {\bibfnamefont {R.}~\bibnamefont {Collins}}, \bibinfo {author} {\bibfnamefont {W.}~\bibnamefont
  {Courtney}}, \bibinfo {author} {\bibfnamefont {A.}~\bibnamefont {Dunsworth}}, \bibinfo {author} {\bibfnamefont {E.}~\bibnamefont {Farhi}}, \bibinfo {author} {\bibfnamefont {B.}~\bibnamefont {Foxen}}, \bibinfo {author} {\bibfnamefont {A.}~\bibnamefont {Fowler}}, \bibinfo {author} {\bibfnamefont {C.}~\bibnamefont {Gidney}}, \bibinfo {author} {\bibfnamefont {M.}~\bibnamefont {Giustina}}, \bibinfo {author} {\bibfnamefont {R.}~\bibnamefont {Graff}}, \bibinfo {author} {\bibfnamefont {K.}~\bibnamefont {Guerin}}, \bibinfo {author} {\bibfnamefont {S.}~\bibnamefont {Habegger}}, \bibinfo {author} {\bibfnamefont {M.~P.}\ \bibnamefont {Harrigan}}, \bibinfo {author} {\bibfnamefont {M.~J.}\ \bibnamefont {Hartmann}}, \bibinfo {author} {\bibfnamefont {A.}~\bibnamefont {Ho}}, \bibinfo {author} {\bibfnamefont {M.}~\bibnamefont {Hoffmann}}, \bibinfo {author} {\bibfnamefont {T.}~\bibnamefont {Huang}}, \bibinfo {author} {\bibfnamefont {T.~S.}\ \bibnamefont {Humble}}, \bibinfo {author} {\bibfnamefont {S.~V.}\ \bibnamefont
  {Isakov}}, \bibinfo {author} {\bibfnamefont {E.}~\bibnamefont {Jeffrey}}, \bibinfo {author} {\bibfnamefont {Z.}~\bibnamefont {Jiang}}, \bibinfo {author} {\bibfnamefont {D.}~\bibnamefont {Kafri}}, \bibinfo {author} {\bibfnamefont {K.}~\bibnamefont {Kechedzhi}}, \bibinfo {author} {\bibfnamefont {J.}~\bibnamefont {Kelly}}, \bibinfo {author} {\bibfnamefont {P.~V.}\ \bibnamefont {Klimov}}, \bibinfo {author} {\bibfnamefont {S.}~\bibnamefont {Knysh}}, \bibinfo {author} {\bibfnamefont {A.}~\bibnamefont {Korotkov}}, \bibinfo {author} {\bibfnamefont {F.}~\bibnamefont {Kostritsa}}, \bibinfo {author} {\bibfnamefont {D.}~\bibnamefont {Landhuis}}, \bibinfo {author} {\bibfnamefont {M.}~\bibnamefont {Lindmark}}, \bibinfo {author} {\bibfnamefont {E.}~\bibnamefont {Lucero}}, \bibinfo {author} {\bibfnamefont {D.}~\bibnamefont {Lyakh}}, \bibinfo {author} {\bibfnamefont {S.}~\bibnamefont {Mandr{\~A} }}, \bibinfo {author} {\bibfnamefont {J.~R.}\ \bibnamefont {McClean}}, \bibinfo {author} {\bibfnamefont {M.}~\bibnamefont
  {McEwen}}, \bibinfo {author} {\bibfnamefont {A.}~\bibnamefont {Megrant}}, \bibinfo {author} {\bibfnamefont {X.}~\bibnamefont {Mi}}, \bibinfo {author} {\bibfnamefont {K.}~\bibnamefont {Michielsen}}, \bibinfo {author} {\bibfnamefont {M.}~\bibnamefont {Mohseni}}, \bibinfo {author} {\bibfnamefont {J.}~\bibnamefont {Mutus}}, \bibinfo {author} {\bibfnamefont {O.}~\bibnamefont {Naaman}}, \bibinfo {author} {\bibfnamefont {M.}~\bibnamefont {Neeley}}, \bibinfo {author} {\bibfnamefont {C.}~\bibnamefont {Neill}}, \bibinfo {author} {\bibfnamefont {M.~Y.}\ \bibnamefont {Niu}}, \bibinfo {author} {\bibfnamefont {E.}~\bibnamefont {Ostby}}, \bibinfo {author} {\bibfnamefont {A.}~\bibnamefont {Petukhov}}, \bibinfo {author} {\bibfnamefont {J.~C.}\ \bibnamefont {Platt}}, \bibinfo {author} {\bibfnamefont {C.}~\bibnamefont {Quintana}}, \bibinfo {author} {\bibfnamefont {E.~G.}\ \bibnamefont {Rieffel}}, \bibinfo {author} {\bibfnamefont {P.}~\bibnamefont {Roushan}}, \bibinfo {author} {\bibfnamefont {N.~C.}\ \bibnamefont {Rubin}},
  \bibinfo {author} {\bibfnamefont {D.}~\bibnamefont {Sank}}, \bibinfo {author} {\bibfnamefont {K.~J.}\ \bibnamefont {Satzinger}}, \bibinfo {author} {\bibfnamefont {V.}~\bibnamefont {Smelyanskiy}}, \bibinfo {author} {\bibfnamefont {K.~J.}\ \bibnamefont {Sung}}, \bibinfo {author} {\bibfnamefont {M.~D.}\ \bibnamefont {Trevithick}}, \bibinfo {author} {\bibfnamefont {A.}~\bibnamefont {Vainsencher}}, \bibinfo {author} {\bibfnamefont {B.}~\bibnamefont {Villalonga}}, \bibinfo {author} {\bibfnamefont {T.}~\bibnamefont {White}}, \bibinfo {author} {\bibfnamefont {Z.~J.}\ \bibnamefont {Yao}}, \bibinfo {author} {\bibfnamefont {P.}~\bibnamefont {Yeh}}, \bibinfo {author} {\bibfnamefont {A.}~\bibnamefont {Zalcman}}, \bibinfo {author} {\bibfnamefont {H.}~\bibnamefont {Neven}},\ and\ \bibinfo {author} {\bibfnamefont {J.~M.}\ \bibnamefont {Martinis}},\ }\bibfield  {title} {\bibinfo {title} {Quantum supremacy using a programmable superconducting processor},\ }\href {https://doi.org/10.1038/s41586-019-1666-5} {\bibfield
  {journal} {\bibinfo  {journal} {Nature}\ }\textbf {\bibinfo {volume} {574}},\ \bibinfo {pages} {505} (\bibinfo {year} {2019})}\BibitemShut {NoStop}%
\bibitem [{\citenamefont {Roffe}(2019)}]{Roffe2019}%
  \BibitemOpen
  \bibfield  {author} {\bibinfo {author} {\bibfnamefont {J.}~\bibnamefont {Roffe}},\ }\bibfield  {title} {\bibinfo {title} {Quantum error correction: an introductory guide},\ }\href {https://doi.org/10.1080/00107514.2019.1667078} {\bibfield  {journal} {\bibinfo  {journal} {Contemporary Physics}\ }\textbf {\bibinfo {volume} {60}},\ \bibinfo {pages} {226} (\bibinfo {year} {2019})}\BibitemShut {NoStop}%
\bibitem [{\citenamefont {Briegel}\ and\ \citenamefont {Raussendorf}(2001)}]{Briegel2001}%
  \BibitemOpen
  \bibfield  {author} {\bibinfo {author} {\bibfnamefont {H.~J.}\ \bibnamefont {Briegel}}\ and\ \bibinfo {author} {\bibfnamefont {R.}~\bibnamefont {Raussendorf}},\ }\bibfield  {title} {\bibinfo {title} {Persistent entanglement in arrays of interacting particles},\ }\href {https://doi.org/10.1103/PhysRevLett.86.910} {\bibfield  {journal} {\bibinfo  {journal} {Phys. Rev. Lett.}\ }\textbf {\bibinfo {volume} {86}},\ \bibinfo {pages} {910} (\bibinfo {year} {2001})}\BibitemShut {NoStop}%
\bibitem [{\citenamefont {Piroli}\ \emph {et~al.}(2021)\citenamefont {Piroli}, \citenamefont {Styliaris},\ and\ \citenamefont {Cirac}}]{Piroli2021}%
  \BibitemOpen
  \bibfield  {author} {\bibinfo {author} {\bibfnamefont {L.}~\bibnamefont {Piroli}}, \bibinfo {author} {\bibfnamefont {G.}~\bibnamefont {Styliaris}},\ and\ \bibinfo {author} {\bibfnamefont {J.~I.}\ \bibnamefont {Cirac}},\ }\bibfield  {title} {\bibinfo {title} {Quantum circuits assisted by local operations and classical communication: Transformations and phases of matter},\ }\href {https://doi.org/10.1103/PhysRevLett.127.220503} {\bibfield  {journal} {\bibinfo  {journal} {Phys. Rev. Lett.}\ }\textbf {\bibinfo {volume} {127}},\ \bibinfo {pages} {220503} (\bibinfo {year} {2021})}\BibitemShut {NoStop}%
\bibitem [{\citenamefont {Tantivasadakarn}\ \emph {et~al.}(2024)\citenamefont {Tantivasadakarn}, \citenamefont {Thorngren}, \citenamefont {Vishwanath},\ and\ \citenamefont {Verresen}}]{Verresen2024}%
  \BibitemOpen
  \bibfield  {author} {\bibinfo {author} {\bibfnamefont {N.}~\bibnamefont {Tantivasadakarn}}, \bibinfo {author} {\bibfnamefont {R.}~\bibnamefont {Thorngren}}, \bibinfo {author} {\bibfnamefont {A.}~\bibnamefont {Vishwanath}},\ and\ \bibinfo {author} {\bibfnamefont {R.}~\bibnamefont {Verresen}},\ }\bibfield  {title} {\bibinfo {title} {Long-range entanglement from measuring symmetry-protected topological phases},\ }\href {https://doi.org/10.1103/PhysRevX.14.021040} {\bibfield  {journal} {\bibinfo  {journal} {Phys. Rev. X}\ }\textbf {\bibinfo {volume} {14}},\ \bibinfo {pages} {021040} (\bibinfo {year} {2024})}\BibitemShut {NoStop}%
\end{thebibliography}%

\end{document}